\documentclass[useAMS,usenatbib,letter,amsmath,fleqn]{mn2e}
\usepackage{graphicx,color}
\usepackage{amssymb,amsmath}
\usepackage{float}
\usepackage{longtable}
\title[Physical models for YORP and Yarkovsky effects]
  {Physical models for the normal YORP and diurnal Yarkovsky effects}
\author[Golubov et al.]
  {O.~Golubov$^{1,2,3}$
  \thanks{E-mail: olexiy.golubov@gmail.com},
  Y. Kravets$^4$, Yu. N. Krugly$^2$, D. J. Scheeres$^3$\\
$^{1}$V. N. Karazin Kharkiv National University, 4 Svobody Sq., Kharkiv, 61022, Ukraine\\
$^{2}$Institute of Astronomy of Kharkiv National University, 35 Sumska Str., Kharkiv, 61022, Ukraine\\
$^{3}$Department of Aerospace Engineering Sciences, University of Colorado at Boulder, 429 UCB, Boulder, CO, 80309, USA\\
$^{4}$Centre de Physique Th\'{e}orique, \'{E}cole Polytechnique, 91128 Palaiseau Cedex, France}
\begin{document}

\date{Accepted ~~. Received ~~; in original ~~}

\pagerange{L\pageref{firstpage}--L\pageref{lastpage}} \pubyear{2015}

\maketitle

\label{firstpage}

\begin{abstract}
We propose an analytic model for the normal YORP and diurnal Yarkovsky effects experienced by a convex asteroid.
Both the YORP torque and the Yarkovsky force are expressed as integrals of a universal function over the surface of an asteroid.
Although in general this function can only be calculated numerically from the solution of the heat conductivity equation,
approximate solutions can be obtained in quadratures for important limiting cases.
We consider 3 such simplified models:
Rubincam's approximation (zero heat conductivity),
low thermal inertia limit (including the next order correction and thus valid for small heat conductivity),
and high thermal inertia limit (valid for large heat conductivity).
All three simplified models are compared with the exact solution.
\end{abstract}

\begin{keywords}
minor planets, asteroids, general
\end{keywords}

\section{Introduction}

The Yarkovsky--O'Keefe--Radzievskii--Paddack (YORP) effect alters the rotation rate of asteroids and the orientations of their rotation axes \citep{rubincam00, bottke06, vokrouhlicky15}.
The vector of the YORP torque has 3 components.
The `axial' component of YORP $T_z$ accelerates or decelerates the asteroid's rotation, changing its angular velocity $\omega$.
The `obliquity' component $T_{\varepsilon}$ turns the rotational axis, changing the obliquity $\varepsilon$.
The third, `precession' component $T_{\Psi}$ causes the precession of the asteroid's axis \citep{bottke06}.
For large asteroids this contribution is negligible compared to gravitational tides that also cause precession \citep{capek04},
however for 100-meter sized asteroids $T_{\Psi}$ can already dominate over the tides.

Initial studies of YORP calculated the effect via integrating the torque over the surface of an asteroid and then averaging it over time
\citep{rubincam00,vokrouhlicky02}, but later it was found more convenient for some applications to change the order of the procedures,
first averaging the force acting on a surface element over time and only then integrating the force over the surface.
\cite{scheeres07} used this approach in his approximate analytic consideration of YORP,
and \cite{breiter09} used it in their numeric simulations of asteroid 25\,143 Itokawa.
\cite{golubov10} constructed in this manner an exact analytic theory for the axial component of YORP.
They performed averaging over time in quadratures, and thus the YORP acceleration was expressed as an integral
over the surface of an asteroid of an analytic function $p_z$ depending on only 2 free parametres, namely the latitude $\psi$ and the obliquity $\varepsilon$.
\cite{golubov10} limited their analysis to the axial component of YORP $T_z$,
but \cite{steinberg11} applied the same methods to get the obliquity component $T_{\varepsilon}$ and the precession component $T_{\Psi}$.
An important drawback of \cite{golubov10} and \cite{steinberg11} was that both conducted their analysis
under the assumption of zero thermal inertia of the surface.
Although this assumption makes no difference for the computation of $T_z$, which is independent of thermal inertia at all,
$T_{\varepsilon}$ strongly depends on thermal inertia \citep{capek04}.

In this article we generalize these results for the case of non-zero thermal inertia.
For a convex asteroid we express $T_{\varepsilon}$ and $T_{\Psi}$ as integrals over the asteroid's surface.
Two universal functions $p_{\sin}$ and $p_{\cos}$, participating in the integrals, incorporate all the necessary information
about the thermal model of the asteroid. In contrast to the function $p_z$, they depend on three parametres, thermal inertia of the surface $\theta$ being the third.

We find functions $p_{\sin}$ and $p_{\cos}$ by numerically solving the heat conductivity equation.
Also we consider limiting cases of very large and very small thermal inertia of the surface,
and develop approximate methods to get solutions in these cases analytically.
The case of zero heat conductivity corresponds to Rubincam's approximation.

In Section \ref{sec:theory} we introduce basic equations describing the YORP and Yarkovsky effects.
We determine the thermal model of the asteroid,
derive general expressions for the YORP and Yarkovsky forces created by a surface element,
and ultimately define the overal YORP torque and the Yarkovsky force as integrals over the surface of the asteroid.
We express the YORP and Yarkovsky forces as integrals of non-dimensionalized temperature,
which is in turn to be determined from a partial differential equation with a non-linear boundary condition.
We call these integrals $p$ with corresponding indices, as physically they mean differently averaged non-dimensionalized pressure.
All these integrals $p$ appear to depend solely on three free parametres, and once parameterized, can be applied to any convex asteroid.

Then in Section \ref{sec:analytic} we propose three simple analytical models to estimate the expressions for $p$ derived in the previous section.
Rubincam's approximation is valid in the case of zero heat conductivity,
the low thermal inertia limit implements a correction for small non-zero heat conductivity,
and the high thermal inertia limit is valid if heat conductivity is very large.

In Section \ref{sec:numeric} we present results of our numeric computations of integrals $p$, study how these depend on all three free parametres,
and compare them with analytical expressions from Section \ref{sec:analytic}.

All these results are obtained under several essential limitations:
the asteroid is assumed convex, its orbit circular, and light scattering Lambertian.
In Section \ref{sec:robustness} we discuss importance of these limitations and possibilities of their surpression.

\section{General theory}
\label{sec:theory}

\subsection{Heat model}

\begin{figure}
\centering
\includegraphics[width=80mm]{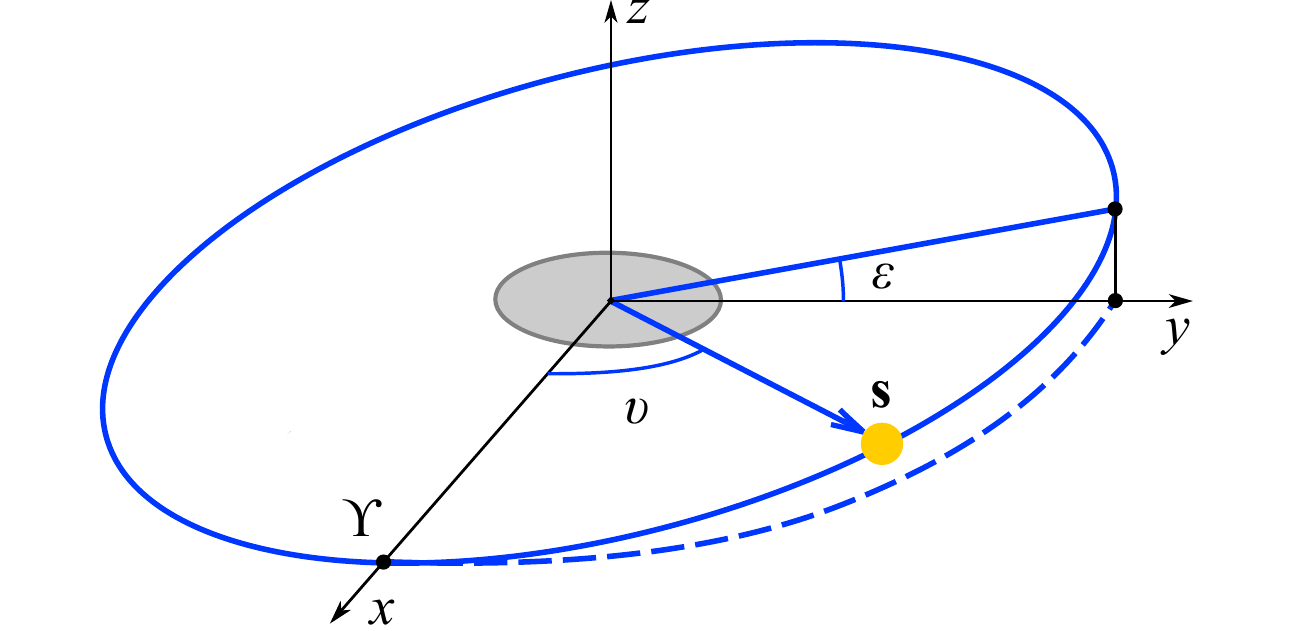}
\caption{Relative position of the asteroid and the Sun.
The coordinate system is centred at the centre of the asteroid,
but does not rotate with it and moves only translationally.
$Oz$ is the rotation axis of the asteroid, $Oxy$ is its equatorial plane.
$\gamma$ stands for the point of the spring equinox of the asteroid.
$\varepsilon$ is the obliquity of the asteroid, the angle between its equatorial and orbital planes.
$\upsilon$ is the angle between the equinox $\gamma$ and the vector $\mathbf{s}$ directed from the asteroid towards the Sun,
and $\upsilon$ corresponds to the time of the year on the asteroid.
In the course of the asteroid's motion around the Sun, the angle $\upsilon$ changes,
and the vector $\mathbf{s}$ circles around the orbit as it is shown in the figure with an ellipsis.}
\label{Fig1}
\end{figure}

\begin{figure}
\centering
\includegraphics[width=80mm]{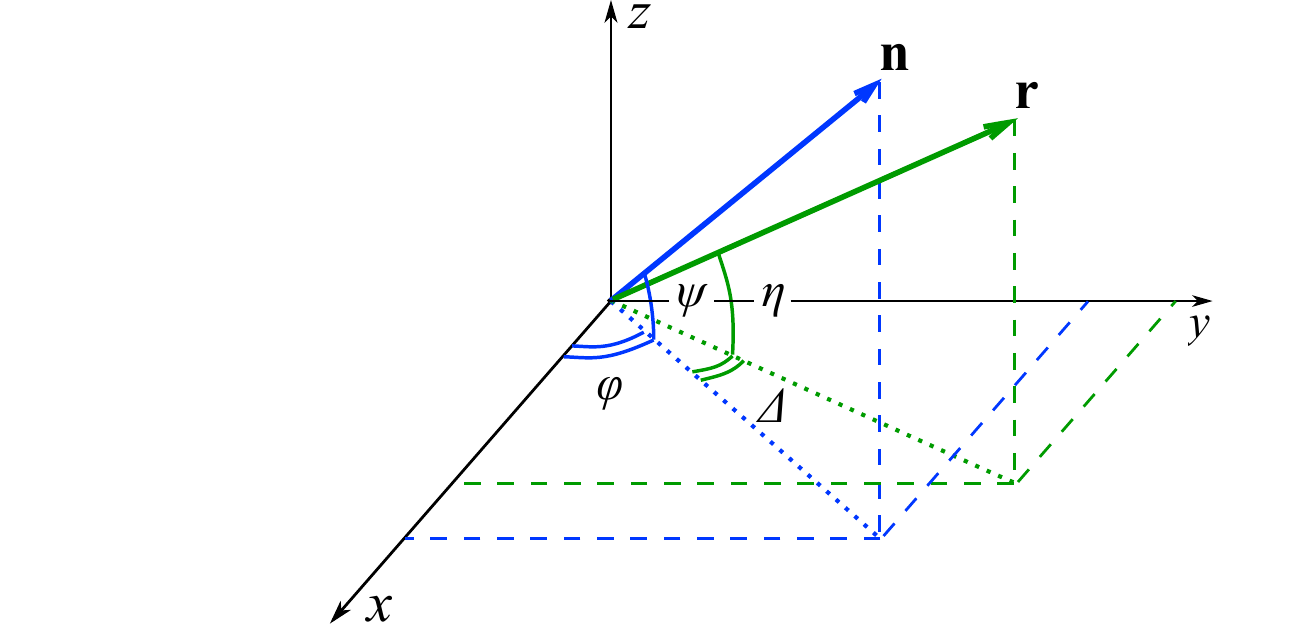}
\caption{Orientation of the normal vector $\mathbf{n}$ of the surface element with respect to the coordinate system.
$\psi$ is latitude of the surface element, determined as the angle between the normal vector $\mathbf{n}$ and the equatorial plane $Oxy$ of the asteroid.
The angle $\phi$ between $Ox$ axis and the projection of $\mathbf{n}$ onto the plane $Oxy$ corresponds to the sidereal time of the surface element.
With the lapse of time the angle $\phi$ changes, and the vector $\mathbf{n}$ rotates around the axis $Oz$.}
\label{Fig2}
\end{figure}

Assuming that on scales essential for heat conductivity the asteroid is large and its surface is flat,
temperature under the surface is governed by the heat conductivity equation in a semispace \citep{lagerros96},
\begin{equation}
\label{conductiv}
\frac{\partial T}{\partial t} = \frac{\kappa}{C\rho} \frac{\partial^2 T}{\partial Z^2}.
\end{equation}
Here $T(Z,t)$ is temperature,
$t$ is time and $Z$ is height above the ground, so that the semi-space under consideration is $Z\le 0$.
Heat conductivity of the material constituting the asteroid surface is $\kappa$,
its heat capacity is $C$,
and its density is $\rho$.

Equation (\ref{conductiv}) requires two boundary conditions and one initial condition.
The boundary condition on the surface of the asteroid is
\begin{equation}
\label{boundary}
\kappa \frac{\partial T}{\partial Z}\bigg|_{Z=0} = E\left(t\right) - \varepsilon\sigma\ T^4\bigg|_{Z=0}.
\end{equation}
The left-hand side and the second term in the right-hand side are related to the surface of the asteroid, $Z=0$.
$E\left(t\right)$ is the radiation flux absorbed by the asteroid's surface.
$\varepsilon\sigma\ T^4\bigg|_{Z=0}$ is the flux irradiated by the surface, with thermal emissivity $\varepsilon$ and the Stefan--Boltzmann constant $\sigma$.
For the second boundary condition, we assume that the asteroid is sufficiently big and flat for the heat flux in its depth to vanish,
\begin{equation}
\label{boundary2}
\kappa \frac{\partial T}{\partial Z}\bigg|_{Z \rightarrow -\infty} = 0.
\end{equation}
Finally, instead of the initial condition we assume periodicity of the solution,
\begin{equation}
\label{boundary3}
T\big|_{t=2\pi/\omega} = T\big|_{t=0}.
\end{equation}
This periodic solution is attained if rotation of the asteroid around its axis is much faster than its rotation around the Sun,
that holds in almost all cases.
Boundary condition in the form of Eqn. (\ref{boundary3}) removes from our consideration such effects as seasonal temperature waves
and variations of day length for eccentric orbits.

The absorbed radiation flux in Eqn. (\ref{boundary}) is determined by the relative orientation of the Sun and the surface element.
Unit vector $\mathbf{s}$ directed towards the Sun is
\begin{equation}
\label{s}
\mathbf{s} = \left(\cos{\upsilon},\ \cos{\varepsilon}\sin{\upsilon},\ \sin{\varepsilon}\sin{\upsilon}\right),
\end{equation}
where $\varepsilon$ is obliquity, i.e. inclination of the asteroid's equator
to its orbit, and $\upsilon$ is the angle between the equinoctial point and the direction towards the Sun (the solar true anomaly).

Normal vector $\mathbf{n}$ of the element $\mathrm{d}S$ is given by
\begin{equation}
\label{n}
\mathbf{n} = \left(\cos{\psi}\cos{\phi},\ \cos{\psi}\sin{\phi},\ \sin{\psi}\right),
\end{equation}
where $\psi$ is the angle between the normal vector $\mathbf{n}$ and the equatorial plane $\left(x, y\right)$,
and $\phi$ is the angle between $x$ axis and projection of $\mathbf{n}$ on the plane $\left(x, y\right)$. $\phi$ can be thought as sidereal time.

According to Eqs. (\ref{s}) and (\ref{n}), cosine of the incidence angle of the radiation $\mathbf{s}\cdot\mathbf{n}$ is given by
\begin{align}
\mathbf{s}\cdot\mathbf{n} = &\cos{\psi}\cos{\phi}\cos{\upsilon} + \nonumber\\
& + \cos{\psi}\sin{\phi}\cos{\varepsilon}\sin{\upsilon} + \sin{\psi}\sin{\varepsilon}\sin{\upsilon}.
\label{sn}
\end{align}
Therefore, the absorbed radiation flux is
\begin{equation}
E\left(t\right) = \left(1 - A\right) \Phi \, \alpha,
\end{equation}
where $A$ is the albedo of the surface, $\Phi$ is solar radiation flux at the asteroid's orbit, and the dimensionless radiation flux $\alpha$ is given by
\begin{align}
\alpha=&(\mathbf{s}\cdot\mathbf{n}) \cdot H(\mathbf{s}\cdot\mathbf{n}),
\label{alpha}
\end{align}
with $H$ standing for the Heaviside step function.

Let us introduce new dimensionless variables. It is appropriate to scale time to the rotation
period, depth to the thermal wave length, and temperature to the equilibrium temperature for subsolar point,
thus introducing
\begin{equation}
t = \frac{\phi}{\omega},\,\,
Z = \sqrt{\frac{\kappa}{C\rho\omega}}\zeta,\,\,
T = \sqrt[4]{\frac{\left(1 - A\right)\Phi}{\varepsilon \sigma}}\tau
\label{tau}
\end{equation}
Then Eqs. (\ref{conductiv})-(\ref{boundary3}) transform into
\begin{equation}
\label{pde}
\frac{\partial \tau}{\partial \phi} = \frac{\partial^2 \tau}{\partial \zeta^2}
\end{equation}
\begin{equation}
\label{boundary_condition}
\theta \frac{\partial \tau}{\partial \zeta}\bigg|_{\zeta=0} = \alpha - \tau^4\bigg|_{\zeta=0}
\end{equation}
\begin{equation}
\label{boundary_condition2}
\frac{\partial \tau}{\partial \zeta}\bigg|_{\zeta \rightarrow -\infty} = 0
\end{equation}
\begin{equation}
\label{boundary_condition3}
\tau\big|_{\phi=2\pi} = \tau\big|_{\phi=0}
\end{equation}
The dimensionless thermal parameter $\theta$ participating in the boundary condition is defined as
\begin{equation}
\theta = \frac{\left(c\rho\omega\kappa\right)^{1/2}}{\left(\varepsilon\sigma\right)^{1/4}\left(1-A\right)^{3/4}\Phi^{3/4}}.
\label{theta}
\end{equation}
It expresses the relative importance of thermal inertia of the surface.
If $\theta \ll 1$, then the left-hand side in Eqn. (\ref{boundary_condition}) is unimportant,
and the surface aquires the equillibrium temperature almost instantly, without any significant thermal lag.
In the opposite case, if $\theta \gg 1$, the thermal lag is so large,
that the surface temperature barely changes at all throughout a rotation period.

It is more convenient to transform Eqn. (\ref{boundary_condition2}) into a different form.
First we integrate Eqn. (\ref{pde}) over $\phi$ from 0 to $2\pi$, and over $\zeta$ from $-\infty$ to 0.
Thus we get
\begin{align}
\int_{-\infty}^0 \left(\tau\big|_{\phi=2\pi}-\tau\big|_{\phi=0}\right)\,\mathrm{d}\zeta=\nonumber\\
\int_{0}^{2\pi} \left(\frac{\partial \tau}{\partial \zeta}\bigg|_{\zeta=-\infty}-\frac{\partial \tau}{\partial \zeta}\bigg|_{\zeta=0}\right)\,\mathrm{d}\phi.
\label{2int_bound_cond}
\end{align}
The left-hand side vanishes because of the periodic initial condition, Eqn. (\ref{boundary_condition3}).
The first term in the right-hand side vanishes because of the boundary condition at $-\infty$, Eqn. (\ref{boundary_condition2}).
Thus Eqn. (\ref{2int_bound_cond}) results into
\begin{equation}
\bigg\langle\frac{\partial \tau}{\partial \zeta}\bigg|_{\zeta=0}\bigg\rangle=0,
\label{average_dtau/dzeta}
\end{equation}
where brackets stand for averaging over $\phi$ from 0 to $2\pi$, i.e. integrating from 0 to $2\pi$ and subsequent division over $2\pi$.
Finally, we average Eqn. (\ref{boundary_condition}) over time, substitute Eqn. (\ref{average_dtau/dzeta}) into its right-hand side, and get
\begin{equation}
\label{boundary_condition_2}
\big\langle\tau^4\big|_{\zeta=0}\big\rangle=\langle{\alpha}\rangle.
\end{equation}
This equation implies that the mean energy emitted by the surface equals the mean energy absorbed by the surface.

\subsection{Instantaneous YORP torque}

The total recoil force $\mathbf{\mathrm{d}f}$ experienced by each surface element consists of three parts:
$\mathbf{\mathrm{d}f}^\mathrm{i}$ produced by the incident solar light,
$\mathbf{\mathrm{d}f}^\mathrm{s}$ produced by solar light scattered by the surface, and
$\mathbf{\mathrm{d}f}^\mathrm{e}$ produced by the infrared light emitted by the heated asteroid's surface.

The first component, $\mathbf{\mathrm{d}f}^\mathrm{i}$, is proved to produce no net YORP torque \citep{rubincam10}.
So further on we disregard the direct solar light pressure $\mathbf{\mathrm{d}f}^\mathrm{i}$.

The expression for the second component depends on the scattering law used.
The simplest and most widely used assumption is Lambert's scattering law.
It assumes no dependence on the direction of the incident light,
and intensity of the emitted light proportional to cosine of the angle with the normal.
Under such assumptions, we get
\begin{equation}
\mathbf{f}^\mathrm{s} = - \frac{2A\Phi}{3c} \alpha \mathbf{\mathrm{d}S}.
\label{f_vis}
\end{equation}
The coefficient 2/3 is caused by Lambert's scattering indicatrix.

The ultimate term, which acts on the surface element due to re-emission of infrared light, is
\begin{equation}
\mathbf{f}^\mathrm{e} = - \frac{2\varepsilon\sigma}{3c} T^4\big|_{Z=0} \mathbf{\mathrm{d}S} = \frac{2(1-A)\Phi}{3c} \tau^4\big|_{\zeta=0} \mathbf{\mathrm{d}S}.
\label{f_ir}
\end{equation}
We again assume Lambert's indicatrix, which results in the same coefficient 2/3.

To compute the YORP torque, take the cross product of the radius-vector $\mathbf{r}$ of the element $\mathrm{d}S$
with the recoil force $\mathbf{\mathrm{d}f}^\mathrm{s}+\mathbf{\mathrm{d}f}^\mathrm{e}$.
The radius-vector is
\begin{equation}
\mathbf{r} = r\left(\cos{\eta}\cos{\left(\phi+\Delta\right)},\ \cos{\eta}\sin{\left(\phi+\Delta\right)},\ \sin{\eta}\right),
\label{r}
\end{equation}
where $\eta$ is the angle between the radius-vector $\mathbf{r}$ and the plane $\left(x, y\right)$,
and $\Delta$ is the angle between projections of $\mathbf{r}$ and $\mathbf{n}$ onto the plane $\left(x, y\right)$.

Thus the torque $\mathbf{\mathrm{d}T}$ created by the recoil force acting upon the surface element $\mathrm{d}S$, is given by the formula
\begin{equation}
\mathbf{\mathrm{d}T} = \frac{2\Phi\mathrm{d}S}{3c} (A\alpha+(1-A)\tau^4\big|_{\zeta=0}) (\mathbf{n} \times \mathbf{r})
\label{dT}
\end{equation}

This torque rapidly changes as the asteroid rotates.
To study the secular evolution we must average $\mathbf{\mathrm{d}T}$ over time.
If spin and orbital periods are non-commensurate, this is equivalent to separate averaging
over spin phase $\phi$ and orbital phase $\upsilon$.
This averaging is carried out in the next two subsections,
first for the axial component $\mathrm{d}T_z$,
and then for the obliquity component $\mathrm{d}T_\varepsilon$
and the precession component $\mathrm{d}T_\Psi$.
Such separation is convenient because of some essential simplifications,
which can be done only for the axial component.

\subsection{Axial component of YORP}
\label{sec:axial}

Substituting Eqs. (\ref{n}) and (\ref{r}) into the $z$-component of Eqn. (\ref{dT}), we get
\begin{equation}
\label{Tz1}
\mathrm{d}T_z = \frac{2\Phi r\mathrm{d}S}{3c} \left( A\alpha+(1-A)\tau^4\big|_{\zeta=0}\right) \sin{\Delta}\cos{\eta}\cos{\psi}.
\end{equation}
It is crucial, that for the $z$-component of torque all terms with $\phi$ and $\upsilon$ cancel,
so that $\mathrm{d}T_z$ depends on time only via the term in brackets.
It drastically simplifies time averaging.
Averaging Eqn. (\ref{Tz1}) and applying boundary condition in the form of Eqn. (\ref{boundary_condition_2}), we get
\begin{equation}
\label{T_z}
\langle{\mathrm{d}T_z}\rangle = \frac{\Phi r\mathrm{d}S}{c} \sin{\Delta}\cos{\eta}\cos{\psi} \, p^\alpha_z,
\end{equation}
where the mean dimensionless pressure $p^\alpha_z$ is determined as
\begin{equation}
\label{p_a_z_def}
p^\alpha_z\left(\psi, \varepsilon\right) = \frac{1}{6\pi^2} \int\limits^{2\pi}_{0} \mathrm{d}v \int\limits^{2\pi}_{0} \mathrm{d}\phi\ \tau^4 \bigg|_{\zeta=0} \equiv \frac{1}{6\pi^2} \int\limits^{2\pi}_{0} \mathrm{d}v \int\limits^{2\pi}_{0} \mathrm{d}\phi\ \alpha
\end{equation}

Changing the order of integration in Eqn. (\ref{p_a_z_def}), and performing the integration over $\upsilon$, we can transform $p^\alpha_z$ to the form
\begin{align}
\label{p_a_z_simpl}
p^\alpha_z\left(\psi, \varepsilon,\theta\right) = &\frac{2}{3\pi^2} \int\limits^{\pi/2}_{-\pi/2} \mathrm{d}\phi \times \nonumber \\
& \times \sqrt{1 - \left(\sin{\phi}\cos{\psi}\sin{\varepsilon} - \sin{\psi}\cos{\varepsilon}\right)^2}\ .
\end{align}

Equations (\ref{p_a_z_def}) and (\ref{p_a_z_simpl}) were previously derived by
\cite{golubov10} and \cite{steinberg11}.

An important and robust by-product of our analysis is that the axial component of YORP
does not depend on the thermal model of the surface, and is determined purely geometrically.
This result is physically very sensible:
for a big locally flat asteroid all energy obtained by a surface element is eventually emitted by the same surface element,
and whatever the time lag for this emission is the lever arm around the rotation axis stays the same,
thus the mean axial torque depends solely on the mean absorbed power.
This does not hold for the obliquity component of YORP,
whose lever arm alters during rotation period of the asteroid, thus the time lag between absorption aand re-emission matters.

The fact that the axial component of YORP is independent of the thermal model was first spotted in numerical simulations by \cite{capek04},
although the authors were hesitant to acknowledge generality of their result.
Later this result was proved theoretically by \cite{scheeres07} in a simplified model of YORP with a constant thermal lag,
by \cite{nesvorny08} in a linearized heat conductivity model,
and by \cite{bbc10} via Fourier decomposition of heat conductivity equation.
Although the latter proof is already general enough, we find our proof more straightforward.
Here we have proved axial YORP independence of the thermal model only for a convex asteroid and Lambert's scattering law.
A more general proof of this fact is given in Appendix \ref{sec:T_z independense}.

\begin{figure*}
\begin{center}
 \includegraphics[width=.49\textwidth]{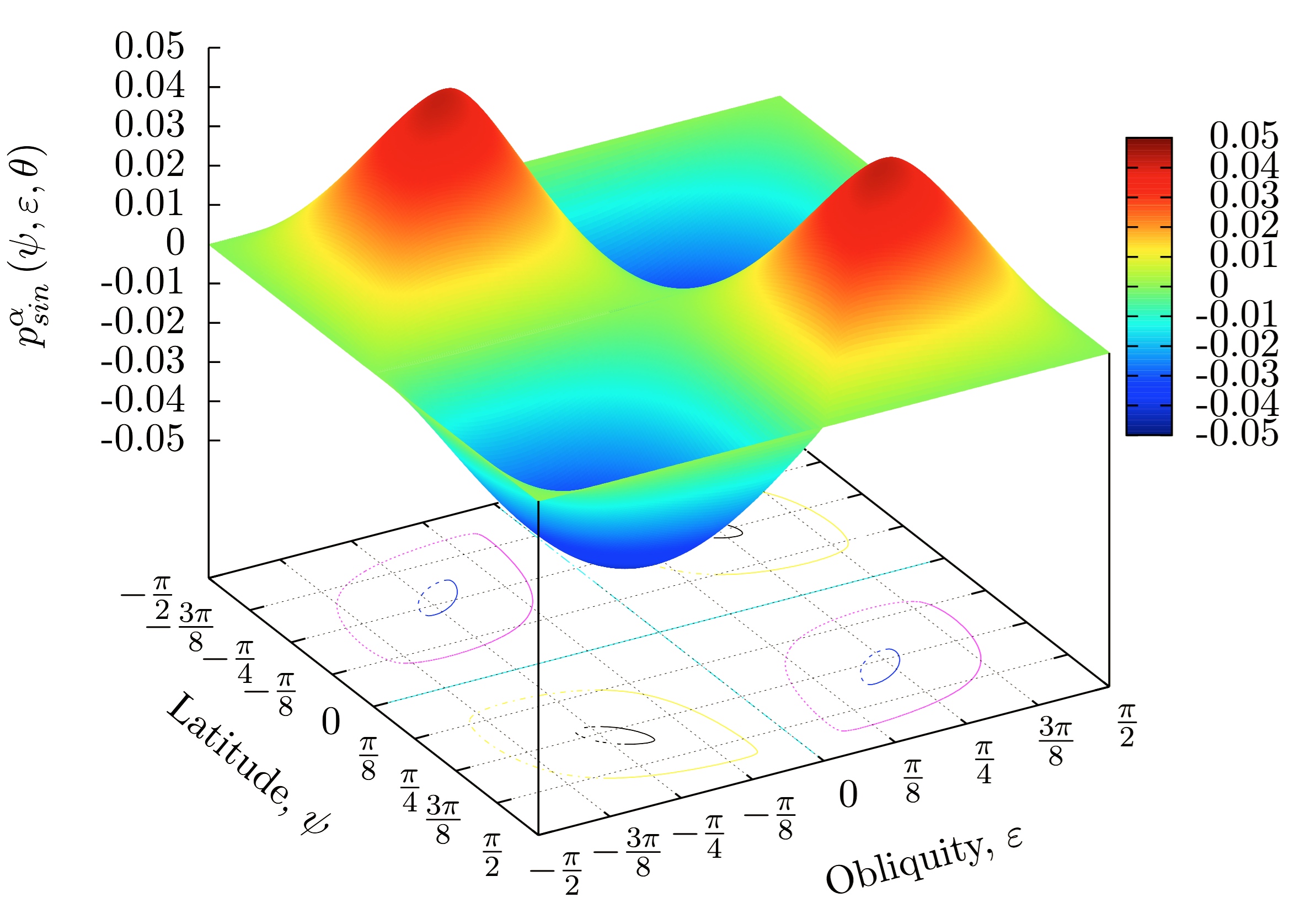}
 \includegraphics[width=.49\textwidth]{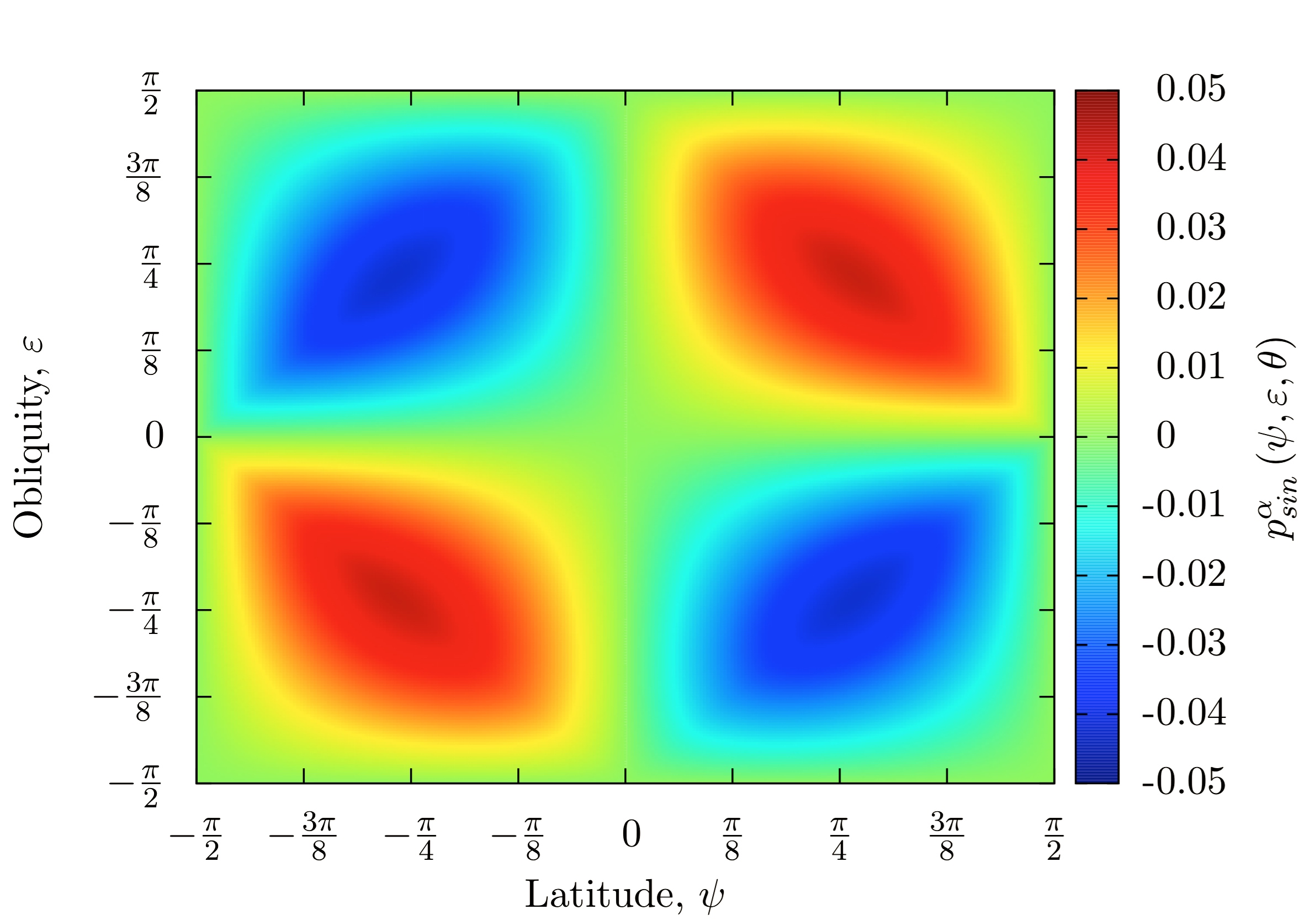}\\
 \includegraphics[width=.49\textwidth]{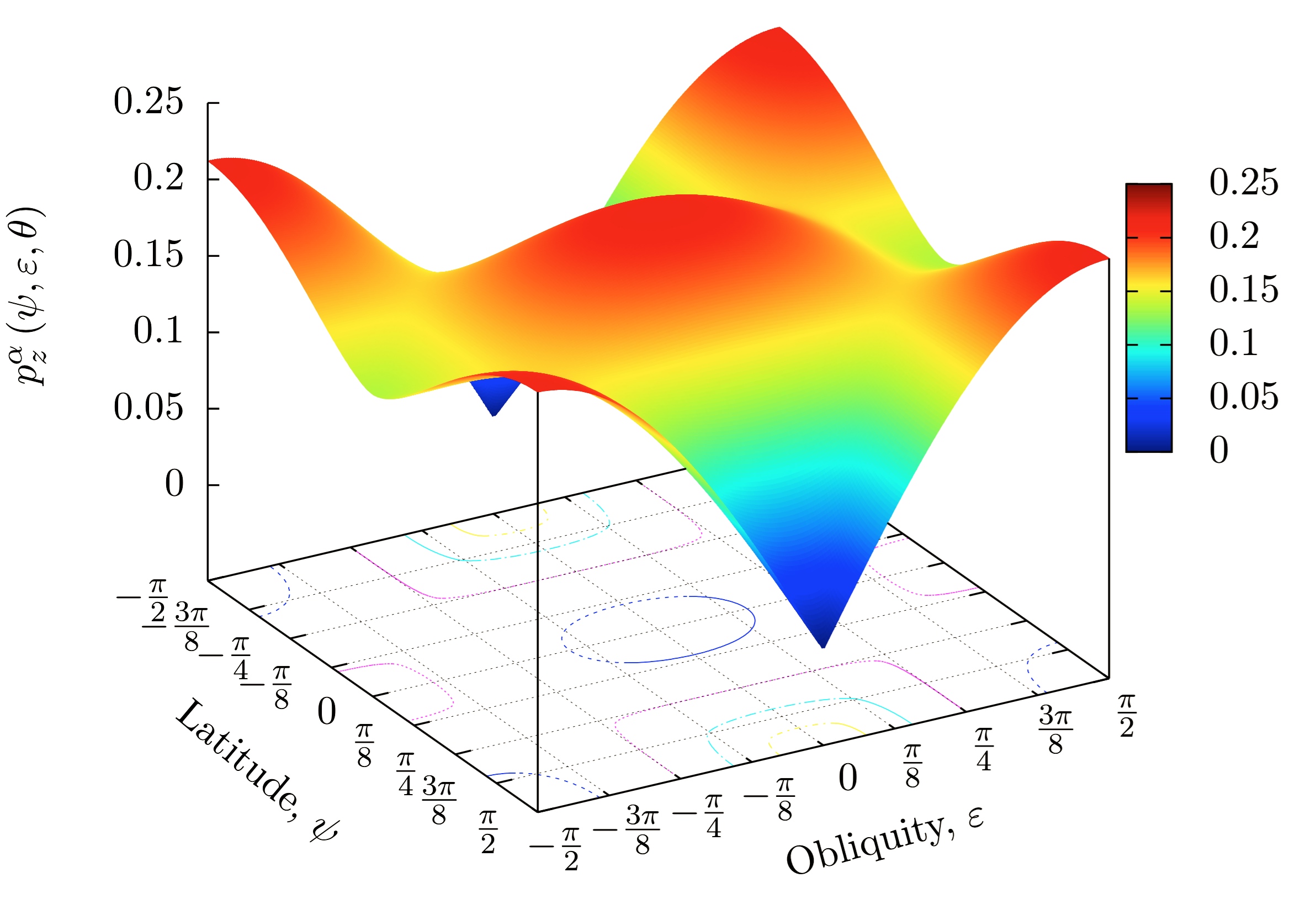}
 \includegraphics[width=.49\textwidth]{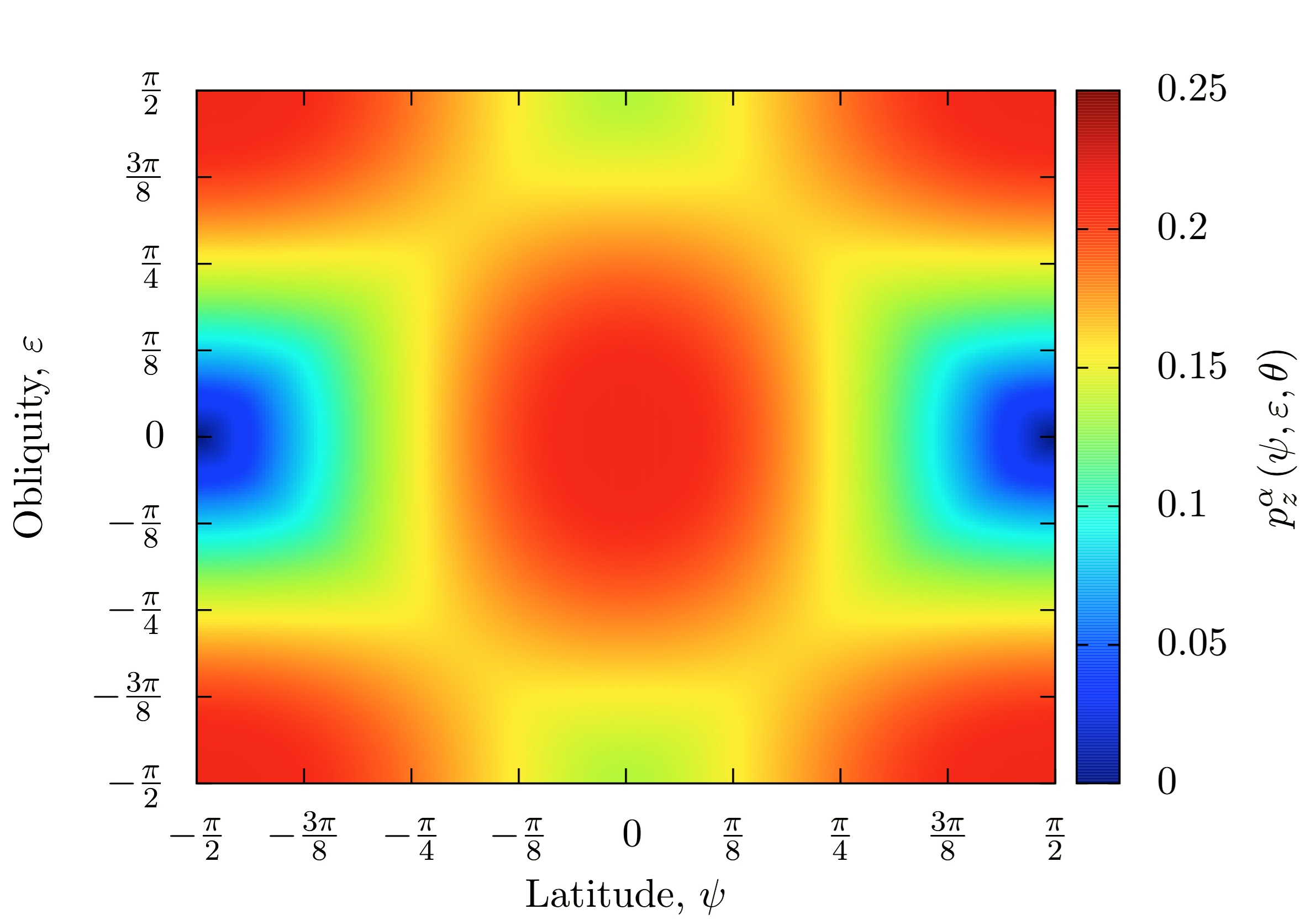}
 \caption{Dimensionless pressures $p_z^\alpha$ and $p_{\sin}^\alpha$ calculated from Eqs. (\ref{p_a_z_simpl}) and (\ref{p_a_sin_simpl}).}
\label{Fig3D}
\end{center}
\end{figure*}

\subsection{Obliquity and precession components of YORP}

Now we substitute Eqs. (\ref{n}) and (\ref{r}) into the $y$ and $x$-components of Eqn. (\ref{dT}), and get
the obliquity component $\mathrm{d}T_{\varepsilon}=\mathrm{d}T_y$ and
the precession component $\mathrm{d}T_{\Psi}=\mathrm{d}T_x$,
\begin{align}
\label{temp1}
\nonumber \mathrm{d}T_{\varepsilon} = &-\frac{2\Phi r\mathrm{d}S}{3c} \left( A\alpha+(1-A)\tau^4\big|_{\zeta=0}\right) \times \\
\nonumber &\times (\sin{\psi} \cos{\eta} \sin{\Delta} \sin{\phi} + \cos{\psi} \sin{\eta} \cos{\phi} -\\
& - \sin{\psi} \cos{\eta} \cos{\Delta} \cos{\phi}) ,
\end{align}
\begin{align}
\nonumber \mathrm{d}T_{\Psi} = &\frac{2\Phi r\mathrm{d}S}{3c} \left( A\alpha+(1-A)\tau^4\big|_{\zeta=0}\right) \times \\
\nonumber &\times (\sin{\psi} \cos{\eta} \cos{\Delta} \sin{\phi} + \sin{\psi} \cos{\eta} \sin{\Delta} \cos{\phi} - \\
&- \cos{\psi} \sin{\eta} \sin{\phi}) .
\end{align}

These equations must be averaged over $\phi$ and $\upsilon$.
This will lead to terms proportional to $\alpha\cos{\phi}$ vanishing,
which can be seen by performing the transformation $\phi \rightarrow \pi - \phi$, \enskip $\upsilon \rightarrow \pi - \upsilon$:
physically it should not affect the result as we still average over the same domain of $\phi$ and $\upsilon$;
but, on the other hand, this transformation changes sign of $\cos{\phi}$ while conserving $\alpha$,
and thus should change sign of $\langle \alpha\cos{\phi} \rangle$.
It is possible only if $\langle \alpha\cos{\phi} \rangle=0$.
Thus after averaging over $\phi$ and $\upsilon$, we get
\begin{align}
\label{T_epsilon}
\langle\mathrm{d}T_{\varepsilon}\rangle = &-\frac{\Phi r \mathrm{d}S}{c} \Big{(}\sin{\psi}\cos{\eta}\sin{\Delta} \times \\
\nonumber &\times \left(Ap^\alpha_\mathrm{sin}\left(\psi, \varepsilon\right)+(1-A)p^\tau_\mathrm{sin}\left(\psi, \varepsilon,\theta\right)\right) + \\
\nonumber & + \left(\cos{\psi} \sin{\eta} - \sin{\psi}\cos{\eta}\cos{\Delta}\right) \times \\
\nonumber & \times (1-A)p^\tau_\mathrm{cos}\left(\psi, \varepsilon,\theta\right)\Big{)},
\end{align}
\begin{align}
\label{T_psi}
\langle\mathrm{d}T_{\Psi}\rangle = &\frac{\Phi r \mathrm{d}S}{c} \Big{(}\sin{\psi}\cos{\eta}\sin{\Delta} \times \\
\nonumber & \times \left(Ap^\alpha_\mathrm{sin}\left(\psi, \varepsilon\right)+(1-A)p^\tau_\mathrm{sin}\left(\psi, \varepsilon,\theta\right)\right) + \\
\nonumber & + \left(\cos{\psi} \sin{\eta} - \sin{\psi}\cos{\eta}\cos{\Delta}\right) \times \\
\nonumber & \times (1-A)p^\tau_\mathrm{cos}\left(\psi, \varepsilon,\theta\right)\Big{)},
\end{align}
where $p^\alpha_\mathrm{sin}$ is determined similarly to Eqn. (\ref{p_a_z_def}),
but with the factor $\sin\phi$ inside the integral,
\begin{equation}
\label{p_a_sin_def}
p^\alpha_\mathrm{sin}\left(\psi, \varepsilon\right) = \frac{1}{6\pi^2} \int\limits^{2\pi}_{0} \mathrm{d}\upsilon \int\limits^{2\pi}_{0} \mathrm{d}\phi\ \alpha \sin{\phi},
\end{equation}
while $p^\tau_\mathrm{sin}$ and $p^\tau_\mathrm{cos}$ are determined through the dimensionless temperature on the surface
\begin{equation}
\label{p_t_sin_def}
p^\tau_\mathrm{sin}\left(\psi, \varepsilon,\theta\right) = \frac{1}{6\pi^2} \int\limits^{2\pi}_{0} \mathrm{d}\upsilon \int\limits^{2\pi}_{0} \mathrm{d}\phi\ \tau^4 \bigg|_{\zeta=0} \sin{\phi},
\end{equation}
\begin{equation}
\label{p_t_cos_def}
p^\tau_\mathrm{cos}\left(\psi, \varepsilon,\theta\right) = \frac{1}{6\pi^2} \int\limits^{2\pi}_{0} \mathrm{d}\upsilon \int\limits^{2\pi}_{0} \mathrm{d}\phi\ \tau^4 \bigg|_{\zeta=0} \cos{\phi}.
\end{equation}
The dimensionless temperature $\tau$ must be defined from Eqn. (\ref{pde}),
which is a partial differential equation with boundary condition Eqn. (\ref{boundary_condition}),
where the expression for $\alpha$ from Eqn. (\ref{alpha}) must be substituted.
One can see that according to these equations $p_\mathrm{sin}$ and $p_\mathrm{cos}$ depend only on 3 parameters:
the latitude $\psi$ of the surface element, the obliquity $\varepsilon$ of the asteroid, and the thermal parameter $\theta$.

Similarly to Eqn. (\ref{p_a_z_simpl}), we can perform integration over $\upsilon$, thus transforming Eqn. (\ref{p_a_sin_def}) into
\begin{align}
\label{p_a_sin_simpl}
p^\alpha_\mathrm{sin}\left(\psi, \varepsilon,\theta\right) = &\frac{2}{3\pi^2} \int\limits^{\pi/2}_{-\pi/2} \mathrm{d}\phi \sin{\phi} \times \\
\nonumber & \times \sqrt{1 - \left(\sin{\phi}\cos{\psi}\sin{\varepsilon} - \sin{\psi}\cos{\varepsilon}\right)^2}\ .
\end{align}

Equation (\ref{p_a_sin_simpl}) is also equivalent to the one derived by \cite{steinberg11},
but Eqs. (\ref{p_t_sin_def}) and (\ref{p_t_cos_def}) depend on the thermal model, which can substantially alter the result.

\subsection{Yarkovsky force}

A similar method can be used to get the diurnal component of the Yarkovsky force created by the surface element.
The force produced by the incident solar light $\mathbf{\mathrm{d}f}^\mathrm{i}$ gives no contribution
to the Yarkovsky force, as the impinging solar photons carry momentum in the radial direction only,
and thus can not change the angular momentum of the asteroid with respect to the Sun.
The remaining two forces $\mathbf{\mathrm{d}f}^\mathrm{s}$ and $\mathbf{\mathrm{d}f}^\mathrm{e}$
given by Eqs. (\ref{f_vis}) and (\ref{f_ir}) respectively must be projected on the direction of the asteroid's orbital motion,
and then averaged over time.
The normal of the orbit is $\mathbf{n}_\mathrm{orb}=(0,-\sin{\varepsilon},\cos{\varepsilon})$.
The direction of the asteroid's motion is $\mathbf{s}\times\mathbf{n}_\mathrm{orb}$.
So we must average the mixed product $(\mathbf{f}^\mathrm{s}+\mathbf{f}^\mathrm{e})\cdot(\mathbf{s}\times\mathbf{n}_\mathrm{orb})$
over $\phi$ and $\upsilon$, thus getting
\begin{align}
\label{F_Yark_def}
\nonumber F_\mathrm{Yark}\left(\psi, \varepsilon,\theta\right) = &\frac{2\Phi\mathrm{d}S}{3c} \big\langle \mathrm{d}\phi (\sin{\psi}\sin{\varepsilon}\cos{\upsilon} + \\
\nonumber & +\cos{\psi}\sin{\phi}\cos{\varepsilon}\cos{\upsilon}-\sin{\upsilon}\cos{\psi}\cos{\phi})\times \\
& \times ((1-A)\tau^4 \big|_{\zeta=0}+A\alpha) \big\rangle.
\end{align}

The term with $\alpha$ corresponding to reflected light vanishes after averaging,
as the transformation $\phi \rightarrow \pi - \phi$, \enskip $\upsilon \rightarrow \pi - \upsilon$ changes its sign.

The term proportional to $\tau^4\cos{\upsilon}$ (product of the first term in the first bracket and the first term in the second bracket) also vanishes.
To prove it we perform the transformation $\phi \rightarrow \pi + \phi - 2\arctan({\cos{\varepsilon}\tan{\upsilon}})$, $\upsilon \rightarrow \pi - \upsilon$.
Its physical meaning is that we are considering the season with the same day length (with the equal number of days before and after solstice),
and the same time of day.
The formula for $\phi$ is complicated because angle $\upsilon$ measured in the orbital plane must be projected onto the equatorial plane.
This transformation is just a translation in $\phi$, so it conserves $\partial /\partial\phi$.
It also conserves $\alpha$, which has a lenthy algebraic proof, but is also pretty evident geometrically.
So Eqs. (\ref{pde}) and (\ref{boundary_condition}) are invariant under this transformation.
It implies that their periodic solution $\tau$ is also conserved.
But $\cos{\upsilon}$ changes its sign, thus the mean of $\tau^4\cos{\upsilon}$ must be equal to its opposite, and therefore must be 0.

After cancelling all zero terms in Eqn. (\ref{F_Yark_def}), we end up with the following expression for the mean Yarkovsky force
\begin{equation}
\langle{\mathrm{d}F_\mathrm{Yark}}\rangle = \frac{\left(1-A\right) \Phi \mathrm{d}S}{c} p^\tau_\mathrm{Yark}\left(\psi, \varepsilon,\theta\right),
\label{F_Yark}
\end{equation}
where the dimensionless Yarkovsky pressure is determined as
\begin{align}
\label{p_Yark_def}
\nonumber p^\tau_\mathrm{Yark} \left(\psi, \varepsilon,\theta\right) = &\frac{1}{6\pi^2} \int\limits^{2\pi}_{0} \mathrm{d}\upsilon \int\limits^{2\pi}_{0} \mathrm{d}\phi \times (\cos{\psi}\sin{\phi}\cos{\varepsilon}\cos{\upsilon} - \\
& - \sin{\upsilon}\cos{\psi}\cos{\phi}) \tau^4 \bigg|_{\zeta=0}.
\end{align}

\subsection{Integrating over the surface}
To get the overal YORP torque and Yarkovsky force acting on an asteroid,
we must integrate Eqs. (\ref{T_z}), (\ref{T_epsilon}), (\ref{T_psi}), (\ref{F_Yark}) over its surface.
Thus we get
\begin{align}
\langle T_z \rangle = \frac{\Phi}{c} \oint\limits_S \mathrm{d}S \,r\sin{\Delta}\cos{\eta}\cos{\psi} \, p^\alpha_z,
\label{T_z_int}
\end{align}
\begin{align}
\langle T_{\varepsilon} \rangle = &-\frac{\Phi}{c} \oint\limits_S \mathrm{d}S \,r\Big{(}\sin{\psi}\cos{\eta}\sin{\Delta} \times \\
\nonumber &\times \left(Ap^\alpha_\mathrm{sin}\left(\psi, \varepsilon\right)+(1-A)p^\tau_\mathrm{sin}\left(\psi, \varepsilon,\theta\right)\right) + \\
\nonumber & + \left(\cos{\psi} \sin{\eta} - \sin{\psi}\cos{\eta}\cos{\Delta}\right) \times \\
\nonumber & \times (1-A)p^\tau_\mathrm{cos}\left(\psi, \varepsilon,\theta\right)\Big{)},
\label{T_epsilon_int}
\end{align}
\begin{align}
\langle T_{\Psi} \rangle = &\frac{\Phi}{c} \oint\limits_S \mathrm{d}S \,r\Big{(}\sin{\psi}\cos{\eta}\sin{\Delta} \times \\
\nonumber &\times \left(Ap^\alpha_\mathrm{sin}\left(\psi, \varepsilon\right)+(1-A)p^\tau_\mathrm{sin}\left(\psi, \varepsilon,\theta\right)\right) + \\
\nonumber & + \left(\cos{\psi} \sin{\eta} - \sin{\psi}\cos{\eta}\cos{\Delta}\right) \times \\
\nonumber & \times (1-A)p^\tau_\mathrm{cos}\left(\psi, \varepsilon,\theta\right)\Big{)},
%\label{T_Psi_int}
\end{align}
\begin{align}
\langle F_\mathrm{Yark} \rangle = \frac{\left(1-A\right) \Phi}{c}\oint\limits_S \mathrm{d}S \, p^\tau_\mathrm{Yark}\left(\psi, \varepsilon,\theta\right),
\label{F_Yark_int}
\end{align}
The dimensionless pressures $p^\alpha_z$, $p^\alpha_\mathrm{sin}$, $p^\tau_\mathrm{sin}$, $p^\tau_\mathrm{cos}$, $p^\tau_\mathrm{Yark}$
must be substituted into these equations.
In Section \ref{sec:analytic} we discuss how these pressures can be estimated analytically,
while in Section \ref{sec:numeric} we compute them numerically.

\section{Approximate analytic solutions}
\label{sec:analytic}
Equations (\ref{T_z_int})-(\ref{F_Yark_int}) express the YORP torque and the Yarkovsky force in terms of several universal functions.
In general case, these functions must be obtained from a numeric solution of the heat conductivity equation.
But in several important general cases they can be expressed in quadratures.

\subsection{Zero thermal inertia}
\label{subsec:zero}
This approximation was first used by \cite{rubincam00}, and now it is often called \textit{Rubincam's approximation}.
It corresponds to instant re-emission of absorbed light by the asteroid's surface.

To obtain Rubincam's approximation from our general approach, we must consider limiting case $\theta \rightarrow 0$.
Then the left-hand side of Eqn. (\ref{boundary_condition}) vanishes,
reducing it to
\begin{equation}
\label{boundary_condition_Rubincam}
\tau \bigg|_{\zeta=0} = \sqrt[4]{\alpha},
\end{equation}
that uniquely determines the surface temperature in the absence of conduction.

We substitute the expression for $\tau$ into Eqs. (\ref{p_t_sin_def}), (\ref{p_t_cos_def}), (\ref{p_Yark_def}),
simplify the results similarly to the previous section, and obtain
\begin{align}
\label{p_sin_R}
p^\tau_\mathrm{sin}\left(\psi, \varepsilon,\theta\right) = &\frac{2}{3\pi^2} \int\limits^{\pi/2}_{-\pi/2} \mathrm{d}\phi \sin{\phi} \times \\
\nonumber & \times \sqrt{1 - \left(\sin{\phi}\cos{\psi}\sin{\varepsilon} - \sin{\psi}\cos{\varepsilon}\right)^2}\ ,
\end{align}
\begin{align}
\label{p_cos_R}
p^\tau_\mathrm{cos}\left(\psi, \varepsilon,\theta\right) = 0,
\end{align}
\begin{align}
\label{p_Yark_R}
p^\tau_\mathrm{Yark}\left(\psi, \varepsilon,\theta\right) = 0.
\end{align}
Thus in Rubincam's approximation $p^\tau_\mathrm{sin}$ is independent of $\theta$,
while $p^\tau_\mathrm{cos}$ and $p^\tau_\mathrm{Yark}$ vanish.
To get non-zero expressions for $p^\tau_\mathrm{cos}$ and $p^\tau_\mathrm{Yark}$,
and also the deviation of $p^\tau_\mathrm{sin}$ from the limit Eqn. (\ref{p_sin_R}),
we must make a correction to Rubincam's approximation.

\subsection{Low thermal inertia}
\label{subsec:low}
Now, instead of substituting $\theta=0$, we will assume $\theta \ll 1$, and treat $\theta$ as a small parameter,
solving Eqs. (\ref{pde}) and (\ref{boundary_condition}) perturbatively.

So, at first we disregard the left-hand side of Eqn. (\ref{boundary_condition}),
and solve Eqn. (\ref{pde}) with the boundary condition Eqn. (\ref{boundary_condition_Rubincam}).
The solution is a Fourier series
\begin{align}
\label{Fourier_Rubincam2}
\nonumber \tau = &\sum\limits^{\infty}_{n=0} A_n \cos\left(n\phi + \sqrt{\frac{n}{2}}\zeta\right)\exp\left(\sqrt{\frac{n}{2}}\zeta\right) + \\
& + \sum\limits^{\infty}_{n=1} B_n \sin\left(n\phi + \sqrt{\frac{n}{2}}\zeta\right)\exp\left(\sqrt{\frac{n}{2}}\zeta\right),
\end{align}
with coefficients
\begin{align}
\nonumber A_0&=\frac{1}{2\pi}\int\limits^{2\pi}_{0}\mathrm{d}\phi\,\sqrt[4]{\alpha},\\
\nonumber A_n&=\frac{1}{\pi}\int\limits^{2\pi}_{0}\mathrm{d}\phi\,\sqrt[4]{\alpha}\cos{n\phi},\\
B_n&=\frac{1}{\pi}\int\limits^{2\pi}_{0}\mathrm{d}\phi\,\sqrt[4]{\alpha}\sin{n\phi}.
\label{coeffients_Rubincam2}
\end{align}

Substituting Eqn. (\ref{Fourier_Rubincam2}) in the left-hand side of Eqn. (\ref{boundary_condition}),
we obtain the corrected expression for the surface temperature.
Then we substitute this temperature into Eqs. (\ref{p_t_sin_def}), (\ref{p_t_cos_def}) and (\ref{p_Yark_def}).
All terms except for the ones with $n=1$ vanish after integration.
Substituting Eqn. (\ref{alpha}) into Eqn. (\ref{coeffients_Rubincam2}), one can see that $A_1=0$.
Therefore $B_1$ is the only Fourier coefficient contributing to the result, and we get
\begin{align}
\label{p_sin_R2}
\nonumber p^\tau_\mathrm{sin}\left(\psi, \varepsilon,\theta\right) = &\frac{1}{6\pi^2} \int\limits^{2\pi}_{0} \mathrm{d}v \int\limits^{2\pi}_{0} \mathrm{d}\phi\,\alpha\sin{\phi} - \\
& - \frac{\theta}{6\sqrt{2}\pi^2} \int\limits^{2\pi}_{0} \mathrm{d}v \int\limits^{2\pi}_{0} \mathrm{d}\phi\,\sqrt[4]{\alpha}\sin{\phi} ,
\end{align}

\begin{align}
\label{p_cos_R2}
p^\tau_\mathrm{cos}\left(\psi, \varepsilon,\theta\right) = - \frac{\theta}{6\sqrt{2}\pi^2} \int\limits^{2\pi}_{0} \mathrm{d}v \int\limits^{2\pi}_{0} \mathrm{d}\phi\,\sqrt[4]{\alpha}\sin{\phi} ,
\end{align}

\begin{align}
\label{p_Yark_R2}
\nonumber  p^\tau_\mathrm{Yark}\left(\psi, \varepsilon,\theta\right) = &\frac{\theta}{6\sqrt{2}\pi^2}
\int\limits^{2\pi}_{0} \mathrm{d}v \int\limits^{2\pi}_{0} \mathrm{d}\phi\,\sqrt[4]{\alpha}\sin{\phi}\times \\
& \times(\cos{\psi}\cos{\varepsilon}\cos{\upsilon}+\sin{\upsilon}\cos{\psi}).
\end{align}

In the limiting case $\theta=0$ Eqs. (\ref{p_sin_R2})-(\ref{p_Yark_R2}) reduce to Eqs. (\ref{p_sin_R})-(\ref{p_Yark_R}).
In Eqn. (\ref{p_sin_R2}) the first-order term derived in this subsection represents a correction to the zero-order term from Rubincam's approximation,
while in Eqs. (\ref{p_cos_R2})-(\ref{p_Yark_R2}) the first-order terms proportional to $\theta$ are the first non-vanishing terms.
Interestingly, the first-order terms in Eqs. (\ref{p_sin_R2}) and (\ref{p_cos_R2}) are the same.

In principle, it is possible to further extend the series in terms of $\theta$.
One might substitute the obtained solution for $\tau$ back into Eqn. (\ref{boundary_condition}),
equate coefficients of different Fourier harmonics, and find the next correction, proportional to $\theta^2$, and so on.
The cost for better accuracy are more complicated expressions.
Equations (\ref{p_sin_R2})-(\ref{p_Yark_R2}) are relatively simple,
but make error of the order of $\theta^2$.

\subsection{High thermal inertia}
\label{subsec:high}

Consider the opposite limiting case of very high thermal inertia, $\theta \gg 1$.
It happens if the asteroid rotates very rapidly, or has very large heat conductivity.
In this case the temperature at each point of the surface varies only slightly during the asteroid's day,
and always remains close to its mean, $\tau_0$.

Then the boundary condition reduces to,
\begin{equation}
\label{boundary_condition_linear}
\theta\frac{\partial \tau}{\partial \zeta}\bigg|_{\zeta=0} = \tau^4_0-\alpha.
\end{equation}
Averaging this equation over $\phi$, and using Eqn. (\ref{average_dtau/dzeta}),
we get the mean temperature $\tau_0$,
\begin{equation}
\tau_0 = \left(\frac{1}{2\pi}\int\limits^{2\pi}_{0} \alpha \mathrm{d}{\phi}\right)^{1/4}.
\end{equation}
This mean temperature varies with orbital phase $\upsilon$.

Solution of Eqn. (\ref{pde}) with periodic boundary condition Eqn. (\ref{boundary_condition3})
is given by Fourier series Eqn. (\ref{Fourier_Rubincam2}).
We substitute this decomposition into boundary condition Eqn. (\ref{boundary_condition_linear}),
equate coeffiecients of corresponding terms, and obtain
\begin{align}
A_0 &= \tau_0^4 ,\nonumber\\
A_n &= \frac{1}{\sqrt{2n}\pi\theta}\int\limits^{2\pi}_{0} \alpha(\cos{n\phi}-\sin{n\phi})\,\mathrm{d}\phi ,\nonumber\\
B_n &= \frac{1}{\sqrt{2n}\pi\theta}\int\limits^{2\pi}_{0} \alpha(\cos{n\phi}+\sin{n\phi})\,\mathrm{d}\phi .
\label{AB_linear}
\end{align}
Comparing the solution given by Eqs. (\ref{Fourier_Rubincam2}) and (\ref{AB_linear}) with the boundary condition Eqn. (\ref{boundary_condition}),
it can be seen that the difference between $\tau$ and $\tau_0$ is indeed of order of $\theta^{-1}$ and thus can be neglected with respect to other terms,
so our assumptions are justified.

Considering that temperature variations are small, we can substitute the term $\tau^4$ in Eqs. (\ref{p_t_sin_def}), (\ref{p_t_cos_def}) and (\ref{p_Yark_def})
with $\tau^4_0 + 4\tau^3_0 (\tau-\tau_0)$.
Only terms proportional to $A_1$ and $B_1$ remain non-zero after integration over $\phi$.
Substituting them from Eqn. (\ref{AB_linear}), we get
\begin{align}
\label{p_sin_linear}
p^\tau_\mathrm{sin}\left(\psi, \varepsilon, \theta\right) = &\frac{\sqrt{2}}{3\pi^2\theta} \nonumber \int\limits^{2\pi}_{0}\mathrm{d}v
\left(\frac{1}{2\pi}\int\limits^{2\pi}_{0}\mathrm{d}\phi\alpha\right)^{3/4} \int\limits^{2\pi}_{0}\mathrm{d}\phi\times \\
& \times \alpha(\cos{\phi}+\sin{\phi}),
\end{align}
\begin{align}
\label{p_cos_linear}
p^\tau_\mathrm{cos}\left(\psi, \varepsilon, \theta\right) = &\frac{\sqrt{2}}{3\pi^2\theta} \nonumber \int\limits^{2\pi}_{0}\mathrm{d}v
\left(\frac{1}{2\pi}\int\limits^{2\pi}_{0}\mathrm{d}\phi\alpha\right)^{3/4} \int\limits^{2\pi}_{0}\mathrm{d}\phi\times \\
& \times \alpha(\cos{\phi}-\sin{\phi}),
\end{align}
\begin{align}
\label{p_Yark_linear}
\nonumber p_\mathrm{Yark}\left(\psi, \varepsilon,\theta\right) = &\frac{\sqrt{2}}{3\pi^2\theta} \int\limits^{2\pi}_{0} \mathrm{d}\upsilon
\left(\frac{1}{2\pi}\int\limits^{2\pi}_{0}\mathrm{d}\phi\alpha\right)^{3/4} \int\limits^{2\pi}_{0}\mathrm{d}\phi \times \\
& \times (\cos{\psi}\cos{\varepsilon}\cos{\upsilon}(\cos{\phi}+\sin{\phi})- \nonumber\\
& - \sin{\upsilon}\cos{\psi}(\cos{\phi}-\sin{\phi})) .
\end{align}

We see that in high thermal inertia limit all dimensionless pressures
$p^\tau_\mathrm{sin}$, $p^\tau_\mathrm{cos}$, and $p^\tau_\mathrm{Yark}$ decay as $\theta^{-1}$.
The following correction proportional to $\theta^{-2}$ might also be obtained
via taking next order term proportional to $\tau-\tau_0$ in the right-hand side of Eqn. (\ref{boundary_condition_linear}),
substituting there the approximate solution, and equating the corresponding terms of the series.
Our approximation disregards this correction, thus error of Eqs. (\ref{p_sin_linear})-(\ref{p_Yark_linear}) is of the order of $\theta^{-2}$.

This approach is more general, than the results by \cite{mysen08} and \cite{nesvorny08},
as \cite{mysen08} limits his consideration to a simplified approximate expression for insolation $\alpha$,
while \cite{nesvorny08} consider only near-spherical shapes of asteroids.
The cost for this generality is a final expression more coplicated, than the ones obtained with additional constraints.

\section{Numerical simulations for the general case}
\label{sec:numeric}

\begin{figure*}
\centering
\includegraphics[width=180mm]{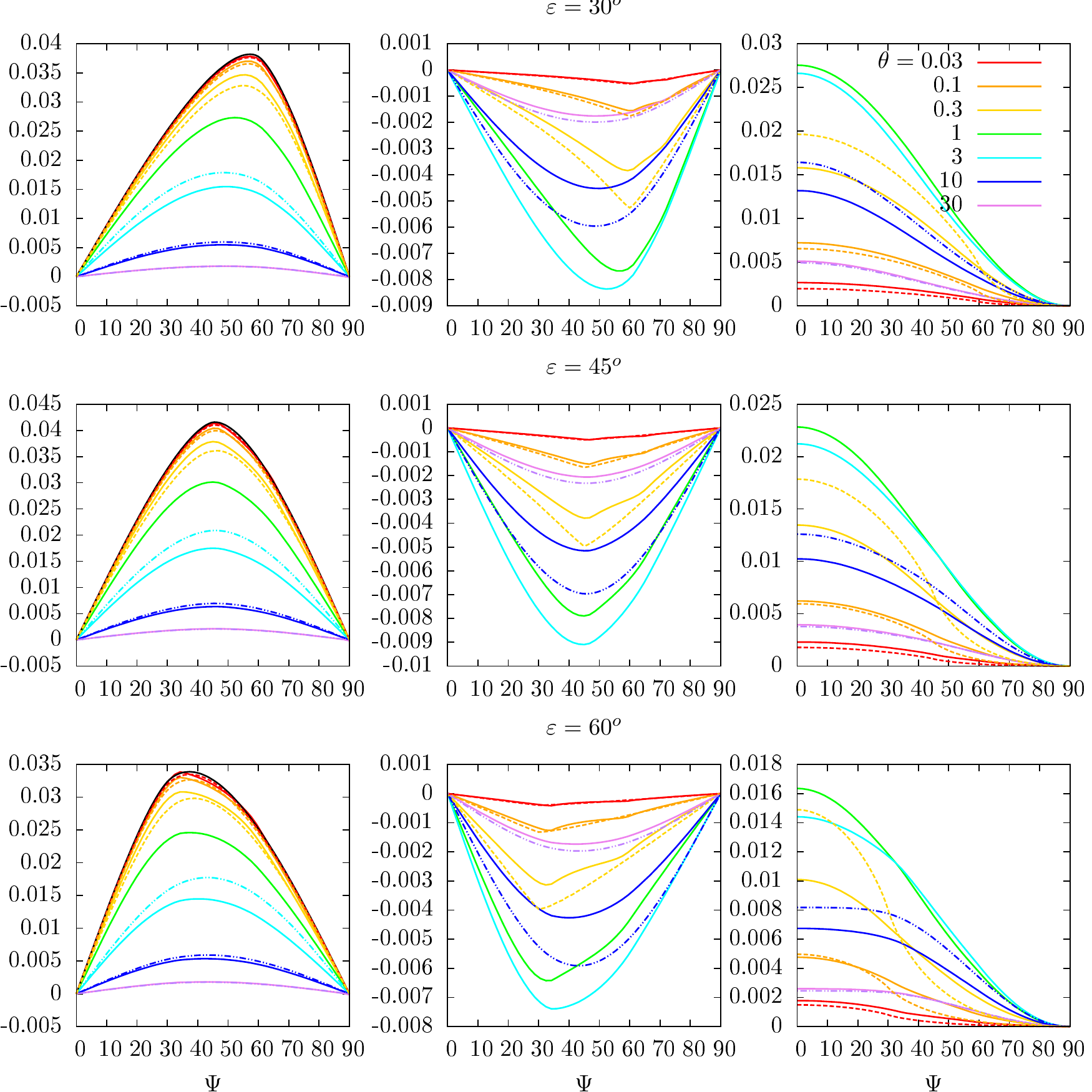}
\caption{Functions $p^\tau_\mathrm{sin}$ (left), $p^\tau_\mathrm{cos}$ (center), and $p^\tau_\mathrm{Yark}$ (right)
for obliquities $\varepsilon=30^{\circ}$ (top), $45^{\circ}$ (center), and $60^{\circ}$ (bottom)
plotted versur latitude $\psi$.
Different colours represent different thermal parametres $\theta$.
Solid coloured lines present results of finite-elements computation (Section \ref{sec:numeric}).
Black line stands for Rubincam's zero thermal inertia approximation (Subsection \ref{subsec:zero}).
Dashed lines show low thermal inertia limit (Subsection \ref{subsec:low}),
dash-dot lines show high thermal inertia limit (Subsection \ref{subsec:high}),
and colour coding is the same as for solid lines.}
\label{p}
\end{figure*}

We created a C++ program, which employed implicit Euler method. We simulate heat conductivity in a horizontal slab to then follow the temperature distribution for several asteroid days giving it enough time to reach the periodic oscillation. Then we compute the YORP torque during the last day. Such computation is repeated for several points evenly distributed along the orbit and the result is averaged. We choose the integration step in time, discretization in space, thickness of the slab, relaxation time, and the number of points on the orbit, such that the result does not change if these parameters are improved. 

Thus our program numerically solves Eqn. (\ref{pde}) with initial and boundary conditions Eqn. (\ref{boundary_condition}--\ref{boundary_condition3})
and calculates the dimensionless pressures from Eqs. (\ref{p_t_sin_def}), (\ref{p_t_cos_def}) and (\ref{p_Yark_def}).
The resulting pressures $p_{\sin}^\tau$, $p_{\cos}^\tau$, and $p_{\mathrm{Yark}}^\tau$ are functions of three parametres $\psi$, $\varepsilon$, and $\theta$.

In Figure \ref{p} we present some cross-sections of these functions.
Here, rows starting from the top one correspond to obliquities $\varepsilon=30^{\circ}$, $45^{\circ}$, and $60^{\circ}$ respectively,
whereas columns from left to right denote dimensionless pressures $p_{\sin}^\tau$, $p_{\cos}^\tau$, and $p_{\mathrm{Yark}}^\tau$.
In each panel we plot corresponding dimensionless pressure versus $\psi$,
for 7 different values of $\theta$ ranging from 0.03 to 30.

Yarkovsky pressure $p_{\mathrm{Yark}}^\tau$ is maximal at $\varepsilon=0$, steadily increases at larger obliquities,
and tends to 0 when $\varepsilon\rightarrow 90^\circ$.
Two pressures corresponding to obliquity component of YORP, $p_{\sin}^\tau$ and $p_{\cos}^\tau$,
vanish at $\varepsilon=0^\circ$ and $\varepsilon=90^\circ$, and attain their maximum inbetween.

As a function of $\psi$, $p_{\mathrm{Yark}}^\tau$ also steadily increases from a maximum at $\psi=0^\circ$ to 0 at $\psi=90^\circ$,
while $p_{\sin}^\tau$ and $p_{\cos}^\tau$ are 0 at $\psi=0^\circ$ and $\psi=90^\circ$, and attain the biggest absolute value at $\psi\approx 45^\circ$.
$p_{\sin}^\tau$ is always positive, while $p_{\cos}^\tau$ is always negative.

With respect to $\theta$, $p_{\sin}^\tau$ steadily decreases, having the maximum at $\theta=0$.
In contrast, $p_{\cos}^\tau$ and $p_{\mathrm{Yark}}^\tau$ tend to 0 when $\theta\rightarrow 0$ and $\theta\rightarrow \infty$,
with an extremum at $\theta\approx 1$.

The black line in the three left panels corresponds to zero heat inertia and is constructed via numeric integration of Eqn. (\ref{p_sin_R}).
Numeric solutions of heat conductivity equation approach this line as $\theta \rightarrow 0$.

Dashed lines are calculated using Eqs. (\ref{p_sin_R2})-(\ref{p_Yark_R2}) in low thermal inertia approximation,
and dash-dot lines are calculated in high thermal inertia approximation with Eqs. (\ref{p_sin_linear})-(\ref{p_Yark_linear}).
Colour coding for these approximate formulae is the same as for exact solutions plotted with solid lines.
Low and high thermal inertia approximations are in good agreement with exact solutions for $\theta \ll 1$ and $\theta \gg 1$ respectively.
The accuracy of 10\% for $p_{\sin}^\tau$ is attained if respectively $\theta < 0.3$ and $\theta > 10$.
For $p_{\cos}^\tau$ and $p_{\mathrm{Yark}}^\tau$ this accuracy is attained if $\theta < 0.1$ and $\theta > 30$.

Some lines demonstrate kinks at $\psi=90^\circ-\varepsilon$.
These kinks are the most prominent in the plot for $p_{\cos}^\tau$ (middle panel) for small $\theta$.
They originate at the polar circle due to appearance and disappearance of polar day and polar night.

\section{Robustness of results}
\label{sec:robustness}
Results obtained in this paper rely on several assumptions, and strictly speaking are applicable only for
convex asteroids on circular orbits with Lambert's scattering and emission indicatrices and locally flat surface.
Now we shall examine these assumptions and discuss their impact on the applicability of our results.

\subsection{Convexity of the asteroid}

\label{sec:non-convex}
\begin{figure*}
\centering
\includegraphics[width=180mm]{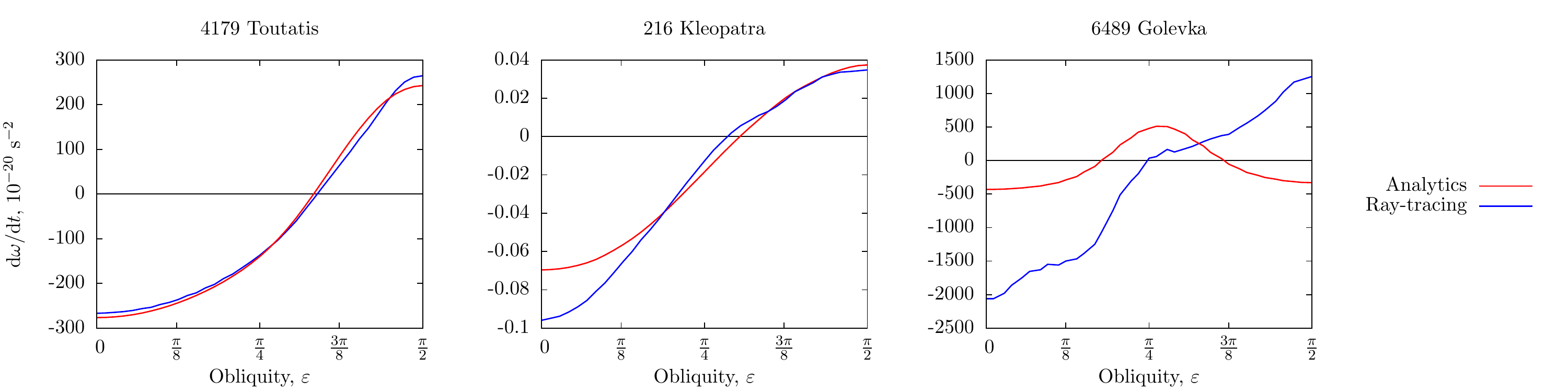}
\caption{YORP accelerations of asteroids 4179 Toutatis, 216 Kleopatra, and 6489 Golevka.
Computations with Eqn. (\ref{T_z}) derived under the assumption of body's convexity,
are compared to ray tracing for non-convex shape.}
\label{ray tracing}
\end{figure*}

Calculation of incoming energy for each surface element is based on the assumption of the asteroid to be convex.
Concave shape leads to two complications: shadowing and self-heating.
However if the asteroid is only moderately concave, shadowing occurs only for rays with glancing angle of incidence,
which anyway contribute little energy per unit area due to small projection onto the normal.
Self-heating for such asteroid can be even less important
because in this case glancing angles of incidence mean not only small projection onto the normal,
but also small emitted intensity in Lambert's law.
Thus we expect that proposed formalism, although developed under the assumption of a convex shape,
will also work reasonably well for moderately concave asteroids.

To test this assumption, we are using a ray tracing program, which calculates the YORP acceleration experienced by non-convex asteroids.
The program randomly casts rays on the surface of the asteroid,
allowing for each of them to be randomly scattered with the directions distributed in accordance with Lambert's law,
tests if it is absorbed by another surface element after the scattering, and if so, it is scattered again, and so on.
The program computes the angular momentum conveyed to the asteroid by the rays.
It assumes instant re-emission of each ray, and thus works in Rubincam's approximation.
We do not separate light into visible and infrared, but assume that Lambert's law describes both scattering and re-emission,
so we just trace a ray of constant energy, and only frequency distribution of energy changes in the ray after scattering/re-emission events.

Plots of the axial component of YORP computed by the program for radar models of three asteroids are shown in Figure \ref{ray tracing}.
Results of the computation using Eqn. (\ref{T_z_int}) are shown for comparison.
We see that moderately non-convex 4179 Toutatis demonstrates a very good agreement between Eqn. (\ref{T_z_int}) and the ray tracing program.
Even for strongly non-convex contact binary 216 Kleopatra the two computations at least qualitatively agree with each other.
And only for extremely non-conxex 6489 Golevka the two curves completely differ.
All other shapes we tested (1620 Geographos, 2063 Bacchus, 1998 KY26) demonstrate good agreement between the two curves, similar to the case with Toutatis.
Thus we conclude, that although Eqs. (\ref{T_z_int})--(\ref{F_Yark_int}) were derived under the assumption of convex shape,
for most real shapes of asteroids they also work reasonably well.

Our simulations are similar to the ones performed by \cite{rozitis13}.
We see a perfect agreement of the shapes of both our analytic and ray tracing plots for 4179 Toutatis and 6489 Golevka
with the corresponding plots in figure 7 in \cite{rozitis13},
although the normalization of the plots is different.

Our program assumes instant re-emission of the absorbed light,
and thus relies on the assumption that Rubincam's approximation can be used for precize calculation of the axial component of YORP.
Although we have proved this theorem only for convex asteroids,
it can be also generalized for the case of concave asteroids.
A proof of this theorem under very general assumptions
is given in Appendix \ref{sec:T_z independense}.

\subsection{Circularity of the orbit}
Throughout the article we assumed the orbit to be circular.
If the orbit is elliptical, the solar energy flux $\Phi$ decreases with the increase of heloicentric distance $r_\odot$.
The angular speed of orbital motion $\mathrm{d}\upsilon/\mathrm{d}t$ also decreases with the increase of $r_\odot$.
In Rubincam's approximation these two factors depending on $r_\odot$ cancel out, and ellipticity of the orbit only results
in a prefactor depending on eccentricity being added to all equations (see Appendix \ref{sec:elliptic} for detail).

Beyond Rubincam's approximation the effect of eccentricity is more substantial.
Still most of our analytic and numerical results can be easily modified to account for eccentricity.
The exact modifications needed are described in Appendix \ref{sec:elliptic}.
Equations do not become much more complicated,
but the number of free parameters increases from 3 to 5,
making the result more difficult to analyze and parameterize.

\subsection{Lambert's law}
We assumed Lambert's law for scattered and emitted light.

If Lambert's law does not hold true for emitted light, but its indicatrix is still independent on the incidence angle,
it causes only a minor correction to our theory.
Indeed, assume we have some emission indicatrix $F^\mathrm{e}(\delta)$,
where $\delta$ is the angle of the emitted ray with the normal to the surface.
Then the recoil pressure created by the emitted light is $P=K^\mathrm{e} \varepsilon \sigma T^4$,
with the coefficient $K^\mathrm{e}$ given by the formula
\begin{equation}
K^\mathrm{e}=\int\limits_0^{\pi/2} 2\pi \cos{i} \sin{i}\, F^\mathrm{e}(\delta) \mathrm{d}\delta.
\end{equation}
For Lambert's scattering law $F^\mathrm{e}(\delta)=\frac{1}{\pi} \cos(\delta)$ we have $K^\mathrm{e}=2/3$,
which we used in Eqn. (\ref{f_vis}).
A different indicatrix will result into coefficient $2/3$ being substituted for the new $K^\mathrm{e}$ in Eqn. (\ref{f_vis})
and all the consequent equations for the force and torque generated by emitted light.

If scattered light is symmetric with respect to the normal, the same consideration also works for it.
However most scattering laws assume that the intensity of scattered light depends not only on the angle $\delta$,
but also on the direction of the incident light.
The same happens if beaming is taken into account for the emitted light \citep{rozitis12}.
The corresponding general theory is developed in Appendix \ref{sec:non-lambertian} and applied to Lommel--Seeliger law.
Mathematically such general theory is not much more complicated than the Lambertian case,
but the formulae are lenghtier and harder to grasp.
Our analysis also shows that corrections due to Lommel--Seeliger law do not substantially change the results.
Therefore Lambert's law for scattered light provides a useful basic approximation.
This agrees with analyzis by \cite{breiter11}, who integrated the torque over the surface of the asteroid
and also found that for Lommel--Seeliger reflection the YORP effect is not very different from the standard Lambertian calculations.

\subsection{Locally flat surface}
Eqs. (\ref{conductiv}) and (\ref{boundary_condition}) assume that heat conductivity is 1-dimensional
and occures in an infinite semispace under the surface.

These assumptions break if the asteroid is small enough for heat fluxes to propagate from one side of the asteroid to the other.
This case has been extensively studied by \cite{breiter10}.

Another important case when heat conductivity may no longer be considered as occuring in 1-dimensional semispace is the case of rough surface,
e.g. stones lying on the surface of an asteroid.
This case analyzed by \cite{golubov12}, \cite{golubov14} and \cite{sevecek15} leads to a concept of tangential YORP,
which appears due to asymmetric emission of light by symmetric stones
and manifests itself in a tangential drag force acting on the surface.

Thus results of this article are applicable only to larger asteroids (at least tens of metres) with relatively smooth surface.

\section{Results}
The main results of this article can be split in three parts.

Firstly, we propose a unified theory for the YORP and Yarkovsky effects,
expressing torques and forces experienced by an asteroid as integrals over its surface.
These integrals include functions, which must be obtained from solutions of heat conductivity equation.
Although these can not be expressed analytically, they depend only on three free parameters,
and once parameterized can be used to calculate the YORP torque and the Yarkovsky force experienced by any asteroid.
This approach is a generalization of results by \cite{golubov10} and \cite{steinberg11}, which were obtained under the assumption of zero heat conductivity.

Secondly, we introduce approximate methods to express these functions in quadratures for the cases of $\theta \ll 1$ and $\theta \gg 1$.
Despite that we only use these approximations to estimate the YORP and Yarkovsky effects,
they can be applied to a much broader range of problems connected with thermal models of asteroidal and planetary surfaces.

Thirdly, we examine the area of applicability of standard treatment of the YORP effect.
We find that moderately concave asteroids still may be safely simulated with our convex model for YORP.
Our model can be generalized for the case of non-Lambertian scattering laws, but the results are very similar to the ones obtained for Lambert's scattering.
Elliptical orbits can also be easily incorporated into our general formalizm.
The independence of the axial component of YORP on the thermal model appears to be very robust,
and holds for concave asteroids and non-Lambertian scattering laws.
This result allows simulating YORP acceleration in Rubincam's approximation.

\section*{Acknowledgements}
We are grateful to the anonymous reviewer, who very much contributed to improving the article.

\appendix

\section{Independence of $T_z$ on the thermal model}
\label{sec:T_z independense}
In Subsection \ref{sec:axial} we have demonstrated that the axial component of YORP $T_z$ is independent of the thermal model.
Although we restricted our proof to convex asteroids only,
it appears that this result does not depend on the convexity of the asteroid
and can be proved under much broader assumptions.

Concavity of the asteroid's shape adds more complications to the proof of applicability of
Rubincam's approximation, but it still appears to hold as long as scattering indicatrix remains independent of the wavelength
and emission indicatrix stays independent of the temperature.
Denoting power coming from surface element $i$ to surface element $j$ at wavelength $\lambda$ via $P_{i\rightarrow j}^{\lambda}$,
we have
\begin{equation}
\label{non-convex1}
P_{i\rightarrow j}^{\lambda} = a_{\odot ij}(t) P_{\odot\rightarrow i}^{\lambda}(t)
+\sum\limits_{k} a_{kij} P_{k\rightarrow i}^{\lambda}
+b_{ij}P_{i\rightarrow}^{\lambda}.
\end{equation}
Here $P_{i\rightarrow}^{\lambda}$ corresponds to the power emitted by element $i$ at wavelength $\lambda$ due to heat radiation,
and $P_{\odot\rightarrow i}^{\lambda}(t)$ is the power coming directly from the Sun to element $i$ at wavelength $\lambda$.
Coefficients $b_{ij}$ here describe the emission indicatrix,
whereas coefficients $a_{kij}$ and $a_{\odot ij}$ the scattering indicatrix.
The first term in the right hand side of Eqn. (\ref{non-convex1}) corresponds to direct sunlight scattered in the direction $i\rightarrow j$,
the sum represents light coming from all other surface elements, which is re-scattered in this direction,
and the final term accounts for emitted light.
Here $a_{\odot ij}(t)$ and $P_{\odot\rightarrow i}^{\lambda}(t)$ explicitly depend on time via position of the Sun.
Coefficients $a$ and $b$ are assumed to be independent of wavelength.

Equation (\ref{non-convex1}) is being integrated over the wavelength and the result being averaged over time.
Denoting all powers integrated over wavelength by the same symbol without supercript $\lambda$,
and assuming chevrons for time averaging,
Eqn. (\ref{non-convex1}) transforms into
\begin{equation}
\label{non-convex2}
\langle P_{i\rightarrow j} \rangle = \langle a_{\odot ij}(t) P_{\odot\rightarrow i}(t) \rangle
+\sum\limits_{k} a_{kij} \langle P_{k\rightarrow i} \rangle
+b_{ij} \langle P_{i\rightarrow} \rangle.
\end{equation}
Heat conservation under each surface element implies that
\begin{equation}
\label{non-convex3}
\langle P_{i\rightarrow} \rangle=\sum\limits_{k}\langle P_{k\rightarrow i} \rangle+\langle P_{\odot\rightarrow i}(t) \rangle.
\end{equation}
Substituting Eqn. (\ref{non-convex3}) into Eqn. (\ref{non-convex2}), we get
\begin{align}
\label{non-convex4}
\langle P_{i\rightarrow j} \rangle
-\sum\limits_{k} a_{kij} \langle P_{k\rightarrow i} \rangle
-b_{ij} \sum\limits_{k}\langle P_{k\rightarrow i} \rangle = \nonumber\\
= \langle a_{\odot ij}(t) P_{\odot\rightarrow i}(t) \rangle
+b_{ij} \langle P_{\odot\rightarrow i}(t) \rangle.
\end{align}
Taken in complex for all $i$ and $j$, Eqn. (\ref{non-convex4}) corresponds to a set of linear equations,
which defines functions $\langle P_{i\rightarrow j} \rangle$, independently of the thermal model.
This leads to Eqn. (\ref{non-convex3}) implying the mean emitted powers $\langle P_{i\rightarrow} \rangle$
to be independent of thermal model.
Therefore the recoil pressures and the YORP effect created by this emission are also constant.
(The latter conclusion also relies on the emission indicatrix being independent of the wavelength.)

Both assumptions of scattering and emission indicatrices being independent of wavelength are essential
for the axial component of YORP to stay independent of the thermal model.
If either of these assumptions breaks down, Eqn. (\ref{non-convex1}) no longer leads to Eqn. (\ref{non-convex2}),
and the heat model enters the YORP effect computations via the wavelength at which heat emission will predominantly occur.

\section{YORP for Elliptical orbits}

\begin{figure*}
\centering
\includegraphics[width=1.\textwidth]{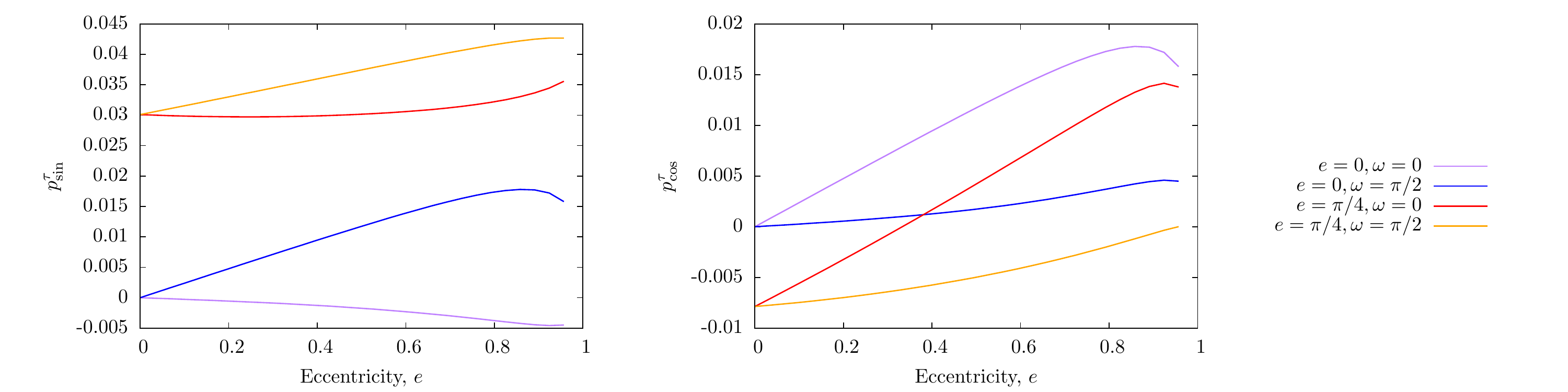}
\caption{Dimensionless pressures $p_{\sin}^{\tau}$ (left) and $p_{\cos}^{\tau}$ (right), as functions of eccentricity.
Different colours represent different obliquities $\varepsilon$ and different arguments of perihelion $\omega$.
For all plots thermal parametre $\theta=1$.}
\label{FigElliptic}
\end{figure*}

\label{sec:elliptic}

If the orbit is elliptical with eccentricity $e\neq0$, we get two important complications in the calculation of YORP.

Firstly, the insolation $\Phi$ is not constant anymore,
but depends on the heliocentric distance $r_\odot$ as
\begin{equation}
\Phi=\frac{\Phi_0}{r_\odot^2},
\label{Phi_elliptic}
\end{equation}
where $\Phi_0$ is solar constant (solar radiation flux at 1 AU),
and $r_\odot$ is measured in astronomic units.

Secondly, angular speed of the asteroid on the orbit varies,
so that orbit average is no longer equal to average over orbital phase $\upsilon$:
time $\mathrm{d}t_\mathrm{orb}$ stands in the same proportion to the area of orbit segment $r_\odot^2\mathrm{d}\upsilon/2$ covered with the elapse of this time,
as the orbital period $T_\mathrm{orb}$ stands to the total area of the orbit $\pi a^2\sqrt{1-e^2}$.
From this proportion we obtain
\begin{equation}
\mathrm{d}t_\mathrm{orb}=\frac{T_\mathrm{orb} r_\odot^2 \mathrm{d}\upsilon}{2\pi a^2\sqrt{1-e^2}}.
\label{dt_elliptic}
\end{equation}

In Rubincam's approximation all our integrals are proportional to $\Phi\mathrm{d}t_\mathrm{orb}$,
thus factors $r_\odot^2$ cancel, and all our analysis for circular orbits is valid,
if only $\Phi$ is substituted with the combination
\begin{equation}
\Phi_1=\frac{\Phi_0}{a^2\sqrt{1-e^2}}.
\label{r_elliptic}
\end{equation}

To go beyond Rubincam's approximation we need to modify our equations more substancially.
First of all, we need the dependence of $r_\odot$ on $\upsilon$, which for elliptic orbit is
\begin{equation}
r_\odot=\frac{a(1-e^2)}{1+e\cos{(\upsilon-\omega)}},
\label{r_elliptic}
\end{equation}
where $\omega$ is argument of perihelion of the asteroid in its own equatorial coordinate frame,
i.e. the angle between the asteroid's spring equinox and its perihelion.

We go through the entire derivation again,
and end up with the same expressions as before,
but with the following substitutions:
\begin{align}
&\alpha \rightarrow \frac{(1+e\cos{(\upsilon-\omega)})^2}{(1-e^2)^{3/2}}\alpha, \label{alpha1}\\
&\mathrm{d}\upsilon \rightarrow \frac{(1-e^2)^{3/2}}{(1+e\cos{(\upsilon-\omega)})^2}\mathrm{d}\upsilon, \label{upsilon1}\\
&\Phi \rightarrow \Phi_1 .
\label{Phi1}
\end{align}
Definitions of $\tau$ and $\theta$ must be also slightly changed by substituting $\Phi_1$ instead of $\Phi$ in Eqs. (\ref{tau}) and (\ref{theta}).

Similarly to Section \ref{sec:analytic} we get expressions
for Rubincam's approximation, low and high thermal inertia limits.
Analogously to Section \ref{sec:numeric} we can simulate Eqs. (\ref{tau}) and (\ref{theta}) with $\alpha$ modified numerically,
and then numerically average it with the modified $\mathrm{d}\upsilon$.

For $p^\alpha_z$ and $p^\alpha_\mathrm{sin}$ additional factors in Eqs. (\ref{alpha1}) and (\ref{upsilon1}) cancel,
and the impact of ellipticity is reduced to alteration of $\Phi$ according to Eqn. (\ref{Phi_elliptic}).

For $p^\tau_\mathrm{sin}$ and $p^\tau_\mathrm{cos}$ the impact of ellipticity is more profound.
We must substitute Eqn. (\ref{alpha1}) into Eqn. (\ref{boundary_condition}), and Eqn. (\ref{upsilon1}) into Eqs. (\ref{p_t_sin_def}) and (\ref{p_t_cos_def}),
thus obtaining corrected expressions for $p^\tau_\mathrm{sin}$ and $p^\tau_\mathrm{cos}$.
Such $p^\tau_\mathrm{sin}$ and $p^\tau_\mathrm{cos}$ are plotted in Figure \ref{FigElliptic}.
They depend not only on $\psi$, $\varepsilon$ and $\theta$, but also on $e$ and $\omega$.
Additional free parameters make these formulae for elliptical orbits harder to paremeterize, and thus less useful.
Still, these formulae provide us with the quantitative estimate of impact of ellipticity of the orbit on the results obtained for circular orbits.

Interestingly, for eccentric orbits $p_{\sin}^{\tau}$ (left) and $p_{\cos}^{\tau}$ can be non-zero even for $\varepsilon=0$.
This behavior agrees with \cite{bbc10}, who predicted non-zero obliquity component of YORP for $\varepsilon=0$ and $e\neq 0$,
and called it \textit{seasonal YORP effect in attitude} by analogy with the seasonal Yarkovsky effect in orbital motion.
Still our analysis proves the analogy relatively weak: the seasonal Yarkovsky effect is caused by seasonal temperature waves,
while the seasonal YORP effect in attitude appears in our model with periodic initial conditions, which exclude seasonal temperature waves.
Thus the simplest possible explanation is that in nonlinear heat conductivity model the mean thermal re-emission lag
is smaller for higher temperatures, so that the torques experienced by the asteroid in different points of its orbit do not compensate each other.
Although the seasonal temperature variations are in turn caused by seasonal temperature waves.

\section{Non-Lambertian scattering laws}
\label{sec:non-lambertian}
\begin{figure}
\centering
\includegraphics[width=80mm]{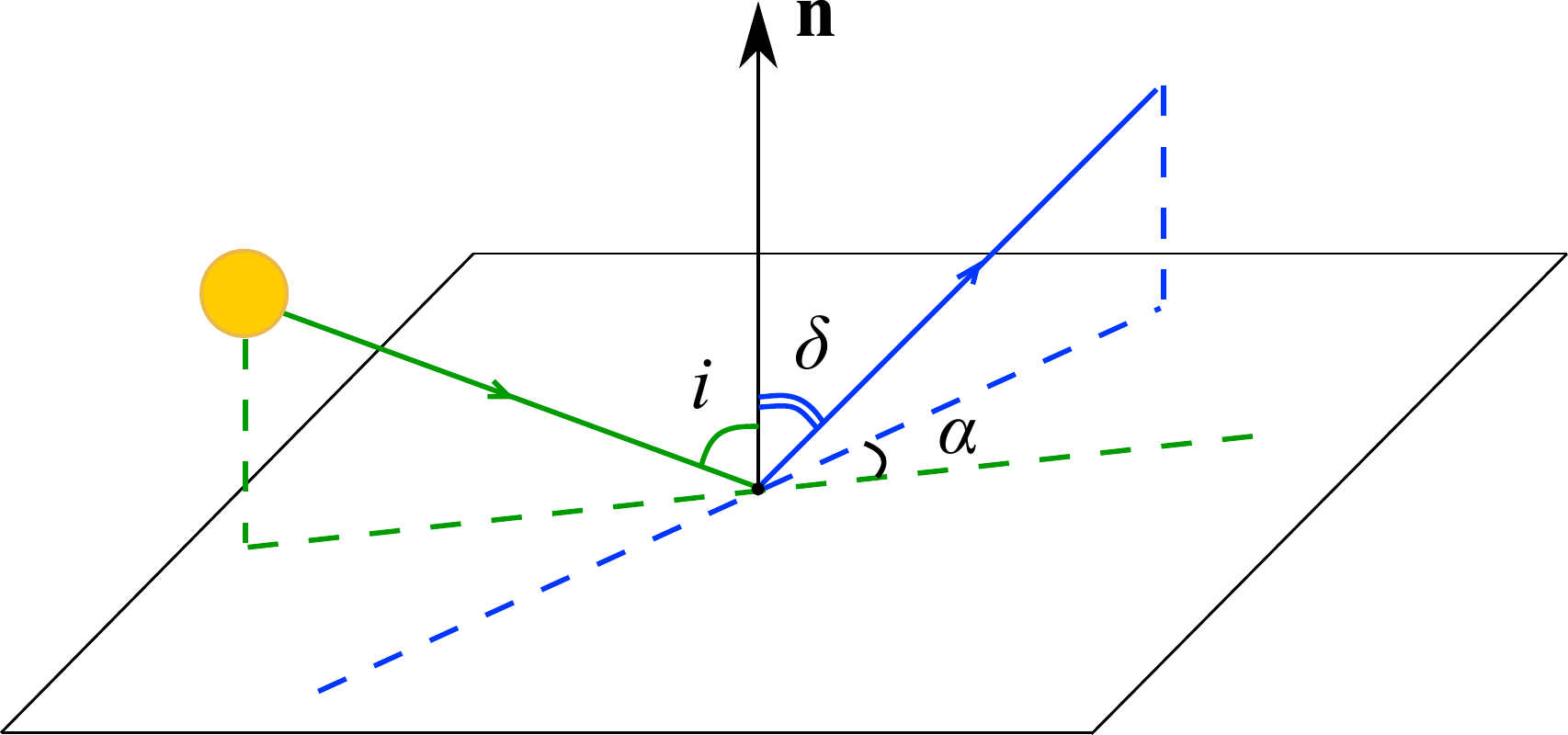}
\caption{Relative position of the incident and the scattered rays.}
\label{Fig9}
\end{figure}
Throughout this paper we assumed Lambert's law to govern scattering of light by the surface of the asteroid.
In this appendix we shall get rid of this assumption and generalize our treatment for any scattering indicatrix.
Let us assume an arbitrary scattering law for visible light,
\begin{equation}
B^\alpha(i,\delta,\alpha)=\Phi \cos{i} F^\alpha(i,\delta,\alpha).
\end{equation}
Here $B^\alpha$ is the brightness of the surface, i.e. energy flux sent by a unit surface element into a unit solid angle.
The angle $i$ between the incident light ray and the normal to the surface can be found from Eqn. (\ref{sn})
by substituting $\cos{i} = \mathbf{s}\cdot \mathbf{n}$ (see Figure \ref{Fig9}).
Also, $\delta$ represents the angle between the emitted light ray and the normal to the surface,
$\alpha$ is the angle between the projections of the incident and the scattered light rays onto the surface of the asteroid,
and $\Phi$ is the incident radiation flux.
The function $F^\alpha$ determines the scattering law.

Let us supplement vector $\mathbf{n}(\cos{\psi} \cos{\phi},$ $\cos{\psi}\sin{\phi},$ $\sin{\psi})$ to an orthogonal basis with vectors
$\mathbf{l}(-\sin{\psi}\cos{\phi}, -\sin{\psi}\sin{\phi}, \cos{\psi})$ and $\mathbf{m}(\sin{\phi}, -\cos{\phi}, 0)$,
directed in the plane of the surface, northward and westward respectively.
Then the recoil force can be expressed in terms of these three basis vectors as follows,
\begin{align}
\mathbf{\mathrm{d}f^\alpha} = & - \frac{\Phi}{c}\cos{i} \,\mathrm{d}S\int\limits^{\pi/2}_0\sin{\delta}\,\mathrm{d}\delta\int\limits^{\pi}_0\mathrm{d}\alpha F^\alpha\left(i,\delta,\alpha\right) \times \\
\nonumber & \left(\mathbf{l}\sin{\delta}\cos{\left(\alpha + \beta\right)} + \mathbf{m}\sin{\delta}\sin{\left(\alpha + \beta\right)} + \mathbf{n}\cos{\delta}\right)\ ,
\end{align}
where $\beta$ is the ``azimuth of the Sun'', i.e. the angle between the north direction $\mathbf{l}$ and the projection of $\mathbf{s}$ onto the horizontal plane,
measured counterclockwise.
It can be determined from the equations
\begin{align}
\nonumber \tan{\beta} = &(- \sin{\psi} \cos{\phi} \cos{v} - \sin{\psi}\sin{\phi}\cos{\varepsilon}\sin{v} + \\
& + \nonumber \cos{\psi}\sin{\varepsilon}\sin{v})/\left(\sin{\phi}\cos{v} - \cos{\phi}\cos{\varepsilon}\sin{v}\right)\ ,
\end{align}
\begin{align}
\mathrm{sign}\left(\cos{\beta}\right) = \mathrm{sign}\left(\sin{\phi}\cos{v} - \cos{\phi}\cos{\varepsilon}\sin{v}\right)\ .
\end{align}

For isotropic surfaces $F^\alpha(i, \delta, \alpha) = F^\alpha(i, \delta, -\alpha)$.
Thus, if we consider two time instances of the same day,
one of which is as close to dawn as the other is to setting ($i_1=i_2$, $\beta_1=-\beta_2$),
and two rays emitted at these instances so that $\delta_1=\delta_2$, $\alpha_1=-\alpha_2$,
we see that the term proportional to $m$ cancels itself after time averaging.
The two resulting terms lead to the following expression for the mean torque created by scattered visible light,
\begin{align}
\label{C4}
\nonumber \langle\mathbf{\mathrm{d}T^\alpha_{z}}\rangle = &\left[\mathbf{r}\times\mathbf{e_z}\right] \mathbf{\mathrm{d}S} \\
\nonumber &\Big\langle\frac{\Phi}{c}\cos{iH}\left(\cos{i}\right)\int\limits^{\pi/2}_0 \sin{\delta}\mathrm{d}\delta \int\limits^{\pi}_0\mathrm{d}\alpha F^\alpha\left(i,\delta,\alpha\right) \times \\
& \times \left(\cos{\delta} - \sin{\delta}\cos{\left(\alpha + \beta\right)}\tan{\psi}\right)\Big\rangle\ .
\end{align}

For the emitted thermal infrared we assume a different indicatrix,
\begin{equation}
B^\tau\left(i,\delta,\alpha\right) = F^\tau\left(\delta\right) P^\tau\ .
\end{equation}

Here $B^\tau$ is brightness of a surface element in infrared,
$P^\tau$ is the total power emitted in infrared,
and $F^\alpha$ is indicatrix, normalized in such a way that
\begin{equation}
\int\limits^{\pi/2}_0 2\pi\sin{\delta}F^\tau\left(\delta\right)\mathrm{d}\delta = 1\ .
\end{equation}

Then the recoil force created by the emitted infrared light is
\begin{align}
\label{C8}
\nonumber \mathbf{\mathrm{d}f^\tau} = & - \frac{\Phi}{c}\cos{i}\mathbf{\mathrm{d}S}K^\tau \\
& \left(1 - \int\limits^{\pi/2}_0 2\pi\sin{\delta}\mathrm{d}\delta\int\limits^\pi_0\mathrm{d}\alpha F^\alpha\left(i,\delta,\alpha\right)\right)\ ,
\end{align}
with $K^\tau$ determined by the expression
\begin{equation}
K^\tau = \int\limits^{\pi/2}_0 2\pi\sin{\delta}\cos{\delta}F^\tau\left(\delta\right)\mathrm{d}\delta\ .
\end{equation}

For example, for Lambert's law $K^\tau = 2/3$, that we have already encountered in Eqn. (\ref{f_vis}).
The force from Eqn. (\ref{C8}) creates the torque
\begin{align}
\label{C9}
\nonumber \langle\mathbf{\mathrm{d}T^\tau_{z}}\rangle = &\left[\mathbf{r}\times\mathbf{e_z}\right]\ \mathbf{\mathrm{d}S}\ \Bigg\langle\frac{\Phi}{c}\cos{iH}\left(\cos{i}\right)K^\tau \\
& \left(1 - \int\limits^{\pi/2}_0 2\pi\sin{\delta}\mathrm{d}\delta\int\limits^\pi_0\mathrm{d}\alpha F^\alpha\left(i,\delta,\alpha\right)\right)\Bigg\rangle\ .
\end{align}

Finally, adding torque from Eqs. (\ref{C4}) and (\ref{C9}) we end up with the same formula Eqn. (\ref{T_z_int})
for the total torque experienced by the asteroid, but with a different expression for the latitude factor.
Instead of Eqn. (\ref{p_a_z_simpl}), now we have
\begin{align}
\label{C10}
\nonumber p(\psi,\varepsilon) = &\int\limits_0^{2\pi}\mathrm{d}\upsilon \int\limits_0^{2\pi}\mathrm{d}\phi \cos{i}H(\cos{i}) \times\\
& \times\bigg(K^\tau+\int\limits_0^{\pi/2}\sin{\delta}\mathrm{d}\delta \int\limits_0^{2\pi}\mathrm{d}\alpha F^\alpha(i, \delta, \alpha)\times\nonumber\\
& \times(\cos{\delta}-\sin{\delta}\cos{(\alpha+\beta)}\tan{\psi}-K^\tau)
\end{align}

If scattering indicatrix depends on the wavelength, then averaging over the entire spectrum must be performed in Eqn. (\ref{C10}).
If emission indicatrix is also different for different wavelengths of the emitted light,
and the ratio between the wavelengths depends on the temperature,
then Rubincam's approximation breaks down as it was discussed at the end of Appendix A,
and our approach can not be used any longer.

Anyway, it appears, that Eqn. (\ref{T_z_int}) has a fairly wide area of applicability, and only the
expression for the latitude factor must be modified for different scattering and emission laws.
Even though the general expression Eqn. (\ref{T_z_int}) is pretty complex, it can be tabulated for any
practically important set of scattering and emission laws, and then used for any shape of an
asteroid.

For instance, we can study the latitude factor for Lommel--Seeliger law,
\begin{equation}
F^\alpha(i, \delta, \alpha)=\frac{C}{\pi} \frac{\cos{\delta}}{\cos{i}+\cos{\delta}}
\end{equation}

\begin{figure}
\centering
\includegraphics[width=80mm]{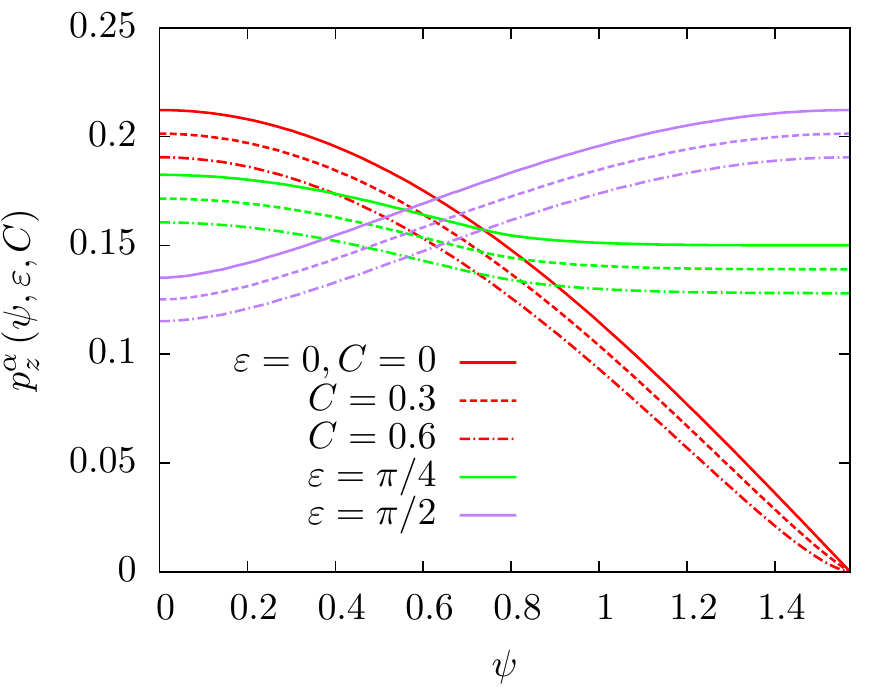}
\caption{Dimensionless pressure $p_{\sin}^\alpha$ calculated from Eqs. (\ref{p_z_non-Lambertian}).}
\label{Fig9}
\end{figure}

Here the constant $C$ roughly corresponds to the albedo of the surface.
This law leads to the following expressions for $p_z^\alpha$ and $p_{\sin}^\alpha$:
\begin{align}
p_z^\alpha(\psi,\varepsilon,C) = &\frac{1}{6\pi^2}\int\limits_0^{2\pi}\mathrm{d}\upsilon \int\limits_0^{2\pi}\mathrm{d}\phi\,\alpha- \nonumber\\
& -\frac{C}{6\pi}\int\limits_0^{2\pi}\mathrm{d}\upsilon \int\limits_0^{2\pi}\mathrm{d}\phi\,\alpha
\bigg(\frac{1}{2}+3\alpha-\nonumber\\
& -\alpha(3\alpha+2) \ln{\bigg(1+\frac{1}{\alpha}\bigg)}\bigg)\bigg|_{\alpha>0}
\label{p_z_non-Lambertian}
\end{align}

This expressions for $p_z^\alpha$ and $p_{\sin}^\alpha$ are plotted against latitude of the surface element in Figure \ref{Fig9}.
Three different curves correspond to different values of the constant $C$: 0, 0.3, and 0.6. In the
case $C = 0$ Eqn. (\ref{p_z_non-Lambertian}) coincides with Eqn. (\ref{p_a_z_simpl}). For bigger $C$ the amount of $f$ is smaller,
but Lambert's law still remains a good approximation.

\label{lastpage}

\end{document}